\documentclass[
reprint,
amsmath,amssymb,
aps,
]{revtex4-2}

\usepackage{xcolor}
\usepackage{graphicx}
\usepackage{dcolumn}
\usepackage{bm}
\usepackage{amsmath,amssymb,amsthm}
\usepackage[normalem]{ulem}
\usepackage{cancel}
\usepackage[nameinlink,capitalize,noabbrev]{cleveref}

\usepackage{tikz}
\usetikzlibrary{arrows.meta,positioning}

\usepackage[most]{tcolorbox}

\newtheoremstyle{ithead-plainbody}
  {6pt}{6pt}{\normalfont}{}{\itshape}{.}{0.5em}{}

\theoremstyle{ithead-plainbody}
\newtheorem{theorem}{Theorem}

\begin{document}

\preprint{APS/123-QED}

\title{Exact and variational identities for free energy differences in strongly coupled open systems}

\author{Mohammad Rahbar}
\affiliation{Technical University of Munich; TUM School of Natural Sciences, Department of Chemistry, Lichtenbergstr. 4, D-85748 Garching, Germany}

\author{Christopher J. Stein}
\affiliation{Technical University of Munich; TUM School of Natural Sciences, Department of Chemistry, Catalysis Research Center, Atomistic Modeling Center, Munich Data Science Institute, Lichtenbergstr. 4, D-85748 Garching, Germany}
\email{christopher.stein@tum.de}

\date{\today}

\begin{abstract}
We derive exact identities for open systems connecting two equilibrium endpoints without imposing microscopic reversibility, detailed balance (DB), fluctuation--dissipation structure, or local detailed balance (LDB) on the driven dynamics.
The identities express the Hamiltonian of mean force (HMF) free energy differences through exponential moments and an explicit chi-squared overlap between the endpoint marginals.
In the frozen-coupling regime, the HMF shift reduces to a bare-system increment and admits a trajectory-level heat--work--reference decomposition.
The exact relations then reduce the problem to a scalar-action law.
A maximum-entropy construction gives a Bessel-form scalar-action law, independent of the microscopic system, environment, and number of degrees of freedom at the level of the variational reconstruction. 
This law provides three outputs from the same sampled configurations: the HMF free energy difference, the endpoint-overlap burden, and a Hessian uncertainty estimate. 
Since many systems in biology, chemistry, physics and engineering violate the underlying assumptions of the standard Jarzynski identity, we validate the framework on a reduced-dimensional model with a non-Liouvillian, phase-space-compressing ramp followed by underdamped Langevin relaxation.
The standard Jarzynski work estimator fails for this ramp because phase-space preservation is broken and no compensating Jacobian correction is included, whereas the present endpoint identities recover the exact HMF free energy difference, and the variational construction reproduces it within its local uncertainty.
\end{abstract}

\maketitle

\begin{figure}[t!]
  \centering
  \includegraphics[width=\columnwidth]{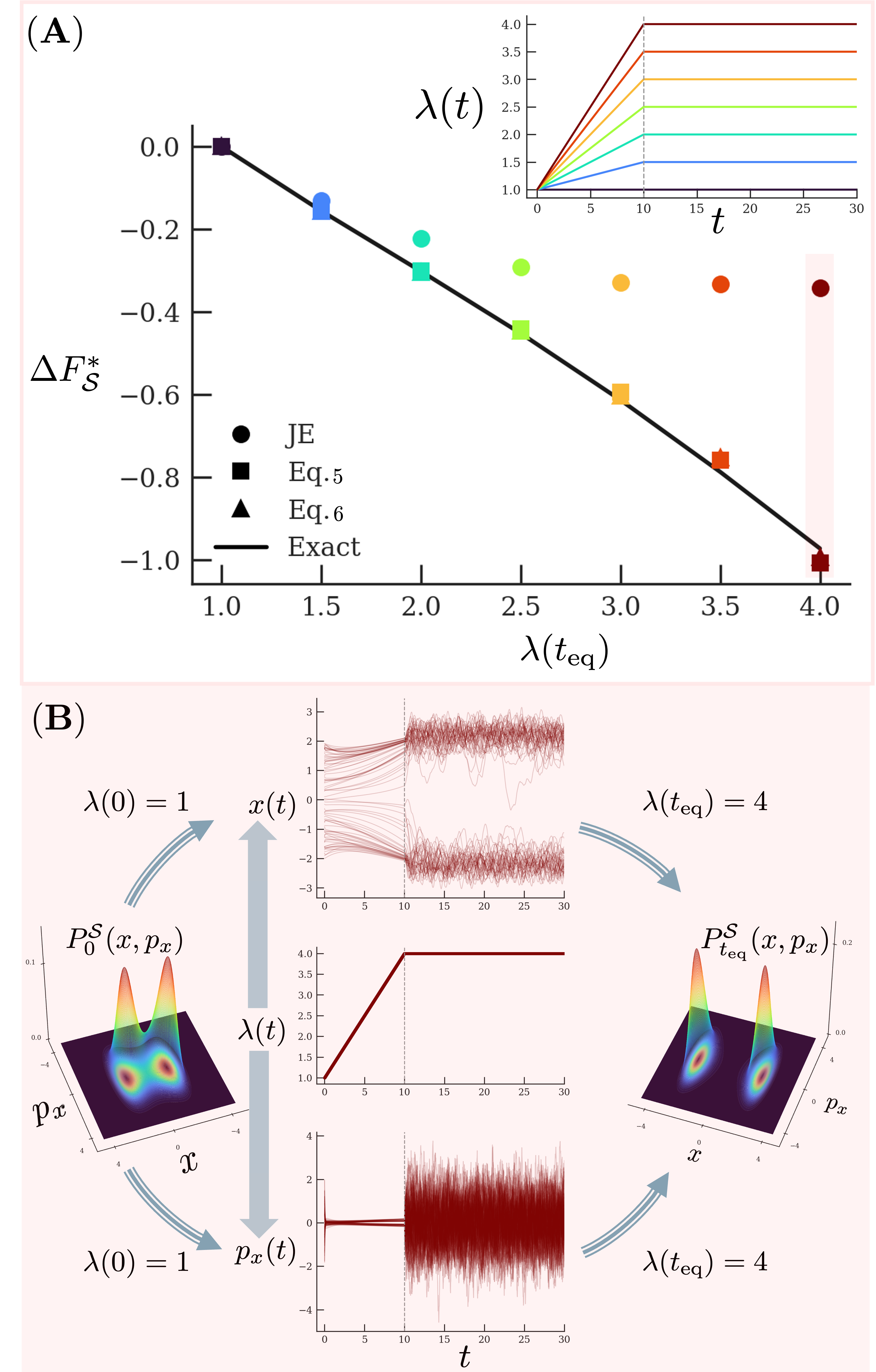}
  \caption{
  \textbf{Validation beyond the JE regime.}
  \textbf{(A)} HMF free energy difference $\Delta F_{\mathcal S}^{*}$ as a function of the final control value $\lambda(t_{\rm eq})$. The solid line is the exact HMF partition-functionreference. Symbols show the JE work estimator, the fixed-coupling ratio identity in Eq.~\eqref{eq_fixc_ratio}, and the complementary moment identity in Eq.~\eqref{eq_fixc_momen}. The JE work estimator fails because the ramp dynamics are non Liouvillian, whereas the present endpoint equalities reproduce the exact endpoint free energy. \textbf{(B)} Representative protocol $\lambda(0)=1\to\lambda(t_{\rm eq})=4$. The center panel shows the ramp and post-ramp relaxation; the dashed line marks the end of the ramp. The upper and lower panels show ensemble trajectories of the system position $x(t)$ and momentum $p_x(t)$. The side panels show the initial and final canonical system marginals, $P_0^{\mathcal S}(x,p_x)$ and $P_{t_{\rm eq}}^{\mathcal S}(x,p_x)$. Post-ramp underdamped Langevin relaxation prepares the final canonical endpoint and satisfies the asymptotic equilibration condition.}
  \label{fig2}
\end{figure}

\noindent
The Jarzynski equality (JE) \cite{jarzynski1997nonequilibrium}, $\langle e^{-\beta W}\rangle=e^{-\beta\Delta F}$, is a fundamental result in modern statistical mechanics.
Here $\beta=(k_BT)^{-1}$ is the inverse temperature, $k_B$ is Boltzmann's constant, $T$ is the temperature, $W$ is work, and $\Delta F$ is the equilibrium free energy difference.
It establishes an exact connection between the exponential average of the nonequilibrium work $W$ performed during an irreversible process and $\Delta F$.
Together with related fluctuation theorems \cite{esposito2009nonequilibrium,crooks1998nonequilibrium,crooks1999entropy}, it has been confirmed in classical and quantum systems and provides a framework for extracting equilibrium information from nonequilibrium measurements
\cite{liphardt2002equilibrium,ohzeki2010quantum,xiong2018experimental, liu2023experimental,harris2007experimental,douarche2005experimental, saira2012test,wimsatt2021harnessing}.
Its generality, however, must be interpreted with care \cite{gore2003bias,toyabe2010experimental,deng2017deformed, hernandez2023experimental,hahn2023quantum,presse2006ordering, chen2008crooks,crooks2009comment,sagawa2010generalized,cohen2004note, cohen2005jarzynski,argun2016experimental,vilar2008failure, palmieri2007jarzynski,monge2018experimental,sone2020quantum}.  
The theoretical foundation of the original JE rests on strict dynamical reversibility \cite{jarzynski1997nonequilibrium}.
In its Hamiltonian derivation, the composite system $\mathcal S+\mathcal E$, with system $\mathcal S$ and environment $\mathcal E$, evolves under deterministic dynamics obeying Liouville's theorem, so phase-space volume is preserved.
A stochastic generalization soon followed, reformulating the dynamics through a Markov master equation or Langevin process \cite{jarzynski1997equilibrium}.
In that setting, reversibility enters statistically through detailed balance (DB), which keeps the Gibbs--Boltzmann distribution stationary for each fixed control parameter $\lambda$. 
The role of these reversibility constraints was sharpened by the criticism of Cohen and Mauzerall \cite{cohen2004note}.
Jarzynski's reply \cite{jarzynski2004nonequilibrium} showed that the JE is a mathematical identity following from Hamiltonian dynamics of the composite system.
Later strong-coupling formulations incorporated the Hamiltonian of mean force (HMF), confirming that the free energy interpretation changes while the proof still relies on Liouville preservation and microscopic reversibility of the composite dynamics \cite{talkner2020colloquium}.
Non-Hamiltonian and non-Markovian extensions then clarified which additional structures can replace Liouville preservation.
Cuendet derived the JE for thermostated non-Hamiltonian Nosé--Hoover dynamics by including the metric factor associated with phase-space compression in the extended phase space \cite{cuendet2006statistical}.
Speck and Seifert extended the equality to generalized Langevin dynamics, provided the friction kernel and noise correlations obey the fluctuation--dissipation theorem (FDT) \cite{speck2007jarzynski}.
Feedback-controlled systems expose another boundary of the standard JE: the equality is restored by adding stochastic mutual information, leading to the generalized Jarzynski equality of Sagawa and Ueda \cite{sagawa2010generalized,toyabe2010experimental}, which still requires local detailed balance (LDB) to define time-reversed path probabilities \cite{maes2021local}.
More generally, Mandal and DeWeese obtained a nonequilibrium work relation for arbitrary non-Hamiltonian dynamics by adding dynamics-dependent terms, including the divergence of the imposed phase-space velocity field, to the exponential weight \cite{mandal2016nonequilibrium}.
Thus, known extensions of the JE beyond Hamiltonian motion either retain a stationary reversibility structure, such as DB or FDT balance, or restore the partition-function ratio by adding a dynamics-specific Jacobian, metric, mutual information, or sink term to the path functional.
The present work takes a complementary route.
We do not seek a modified work functional that repairs a non-Hamiltonian path measure. The present theory is therefore not a generalized work theorem.
Instead, we use the two canonical endpoint marginals as the primary objects and ask what exact identities follow once the driven process relaxes to the final canonical state.
Many driven systems in biology, chemistry, physics, and engineering violate the dynamical assumptions behind standard fluctuation relations while still relaxing to a genuine equilibrium once the driving is held fixed \cite{gaspard2020active,martinez2019inferring,gnesotto2018broken, goswami2019work,godec2022challenges,krishnamurthy2016micrometre, reimann2002brownian,bechinger2016active,caprini2019activity, vicsek2012collective,cates2015motility,fodor2016far,argun2016non}.
This raises the question addressed here: can one recover the free energy difference between two equilibrium endpoints without constructing a dynamics-specific Jacobian correction and without imposing the usual dynamical constraints on the path connecting them? A detailed formulation for arbitrary protocols $(\lambda(t),C(t))$ is given in Ref.~\cite{rahbar2025exact}.
Here $C(t)$ is the coupling control.
We focus on the frozen-coupling regime, $C(t)\equiv C(0)$, which is the standard work-protocol setting: $\lambda(t)$ drives the system while the coupling to the environment remains fixed.
This regime corresponds to the case in which the HMF shift reduces to a bare-system increment and can be written directly in terms of trajectory observables \cite{jarzynski2004nonequilibrium,talkner2020colloquium}.
We consider two canonical endpoints of the composite $\mathcal S+\mathcal E$ at inverse temperature $\beta$.
The reduced equilibrium state of $\mathcal S$ has the HMF form
\begin{equation}
\label{} P^{\mathcal S}(X_{\mathcal S};\lambda,C,\beta)= \frac{e^{-\beta\mathcal H_{\beta}^{*}(X_{\mathcal S},\lambda,C)}}{\mathcal Z_{\mathcal S}^{*}(\lambda,C,\beta)},
\end{equation}
where $X_{\mathcal S}$ is the system phase point, $\mathcal H_{\beta}^{*}$ is the HMF, and $\mathcal Z_{\mathcal S}^{*}$ is the HMF partition function.
The associated HMF free energy is $F_{\mathcal S}^{*}(\lambda,C,\beta) =-\beta^{-1}\ln\mathcal Z_{\mathcal S}^{*}(\lambda,C,\beta)$.
For fixed coupling, the endpoint free energy difference is $\Delta F_{\mathcal S}^{*}(\beta) = F_{\mathcal S}^{*}(\lambda(t_{\rm eq}),C(0),\beta) - F_{\mathcal S}^{*}(\lambda(0),C(0),\beta)$,  where $t_{\rm eq}$ denotes relaxation to the final canonical endpoint after the driving has stopped.
The corresponding endpoint marginals are denoted by $P_0^{\mathcal S}$ and $P_{t_{\rm eq}}^{\mathcal S}$. 
For fixed coupling, the HMF contribution from the environment is independent of $\lambda$ and cancels between the two endpoints.
The endpoint HMF shift therefore reduces to the bare-system increment $\Delta \mathcal H_{\mathcal S}(X_{\mathcal S}) = \mathcal H_{\mathcal S}(X_{\mathcal S},\lambda(t_{\rm eq})) - \mathcal H_{\mathcal S}(X_{\mathcal S},\lambda(0))$, where $\mathcal H_{\mathcal S}$ is the bare-system Hamiltonian.
The endpoint equalities then read
\begin{align}
\label{eq_fix_c_rat_end}
e^{-\beta \Delta F_{\mathcal S}^{*}(\beta)}&=\frac{\left\langle
e^{-\beta\Delta \mathcal H_{\mathcal S}(X_{\mathcal S})}\right\rangle_{\mathcal S}}{1+\chi^2\!\left(P^{\mathcal S}_{t_{\rm eq}}\parallel P^{\mathcal S}_{0}\right)},
\\[3pt]
\label{eq:endpoint_fixedC_moment}
e^{+\beta \Delta F_{\mathcal S}^{*}(\beta)}
&=
\left\langle
e^{+\beta\Delta \mathcal H_{\mathcal S}(X_{\mathcal S})}
\right\rangle_{\mathcal S},
\end{align}
where $\langle\bullet\rangle_{\mathcal S}$ denotes averaging over $P^{\mathcal S}_{t_{\rm eq}}$. Here $\chi^2(P^{\mathcal S}_{t_{\rm eq}}\parallel P^{\mathcal S}_{0})$ denotes the chi-squared divergence of the final endpoint marginal $P^{\mathcal S}_{t_{\rm eq}}$ relative to the initial endpoint marginal $P^{\mathcal S}_{0}$,
\begin{equation}
\chi^2(P^{\mathcal S}_{t_{\rm eq}}\parallel P^{\mathcal S}_{0})
=
-1+
\int \mathrm dX_{\mathcal S}\,
\frac{
\left[P^{\mathcal S}_{t_{\rm eq}}(X_{\mathcal S},\beta)\right]^2
}{
P^{\mathcal S}_{0}(X_{\mathcal S},\beta)
}.  
\end{equation}
Equations~\eqref{eq_fix_c_rat_end} and \eqref{eq:endpoint_fixedC_moment} are the fixed-coupling reduction of the general HMF endpoint identity derived in Ref.~\cite{rahbar2025exact}; the derivation is given in the Appendix . If $C(t)$ is not fixed, the cancellation no longer occurs and the full HMF increment must be retained.
The endpoint identities depend only on the two canonical marginals; they impose no microscopic reversibility, DB, FDT relation, or LDB on the dynamics that connects them.
To pass from endpoints to trajectories, let $\mathcal T_t^{\mathcal S}$ be the system projection of the driven map.
We impose asymptotic equilibration $\lim_{t\to t_{\rm eq}} P_t^{\mathcal S}(X_{\mathcal S},\beta) = P^{\mathcal S}_{t_{\rm eq}}(X_{\mathcal S},\beta)$ as an endpoint boundary condition, not as a reversibility constraint.
This pushforward step converts Eqs.~\eqref{eq_fix_c_rat_end} and \eqref{eq:endpoint_fixedC_moment} into
\begin{align}
\label{eq_fixc_ratio}
e^{-\beta \Delta F_{\mathcal S}^{*}(\beta)}
&=
\frac{
\left\langle
e^{-\beta\Delta \mathcal H_{\mathcal S}(X_0)}
\right\rangle_{X_0}
}{
1+\chi^2\!\left(P^{\mathcal S}_{t_{\rm eq}}\parallel P^{\mathcal S}_{0}\right)
},
\\[3pt]
\label{eq_fixc_momen}
e^{+\beta \Delta F_{\mathcal S}^{*}(\beta)}
&=
\left\langle
e^{+\beta\Delta \mathcal H_{\mathcal S}(X_0)}
\right\rangle_{X_0},
\end{align}
where $\langle\bullet\rangle_{X_0}$ denotes averaging over the initial canonical composite ensemble, $X_0$ is an initial composite phase point, and $\Delta \mathcal H_{\mathcal S}(X_0) = \mathcal H_{\mathcal S} (\mathcal T_{t_{\rm eq}}^{\mathcal S}(X_0),\lambda(t_{\rm eq})) - \mathcal H_{\mathcal S} (\mathcal T_{t_{\rm eq}}^{\mathcal S}(X_0),\lambda(0))$.
The derivation is given in the Appendix.
Thus, the trajectory form is exact once the propagated ensemble has reached the final canonical HMF marginal; no reversibility condition is imposed on the driven part of the process.
For fixed coupling, the same trajectory increment admits the pathwise heat--work--reference decomposition
\begin{equation}
\label{eq_heat_work_ref}
\Delta \mathcal H_{\mathcal S}(X_0)
=
W_{\mathcal S}(t_{\rm eq}|X_0)
+
Q_{\mathcal S}(t_{\rm eq}|X_0)
-
II(t_{\rm eq}|X_0).
\end{equation}
Here $W_{\mathcal S}$ is the work-like contribution, $Q_{\mathcal S}$ is the heat-like contribution, and $II$ is a reference projection.
Their definitions, together with the corresponding trajectory representations of the endpoint identities, the relation to the generalized JE, and the recovery of the JE as a limiting case are given in the End Matter and in the Appendix. Fig.~\ref{fig2} validates these identities for a non-Liouvillian ramp followed by canonical relaxation.
 
\begin{figure}[t!]
  \centering
  \includegraphics[width=\columnwidth]{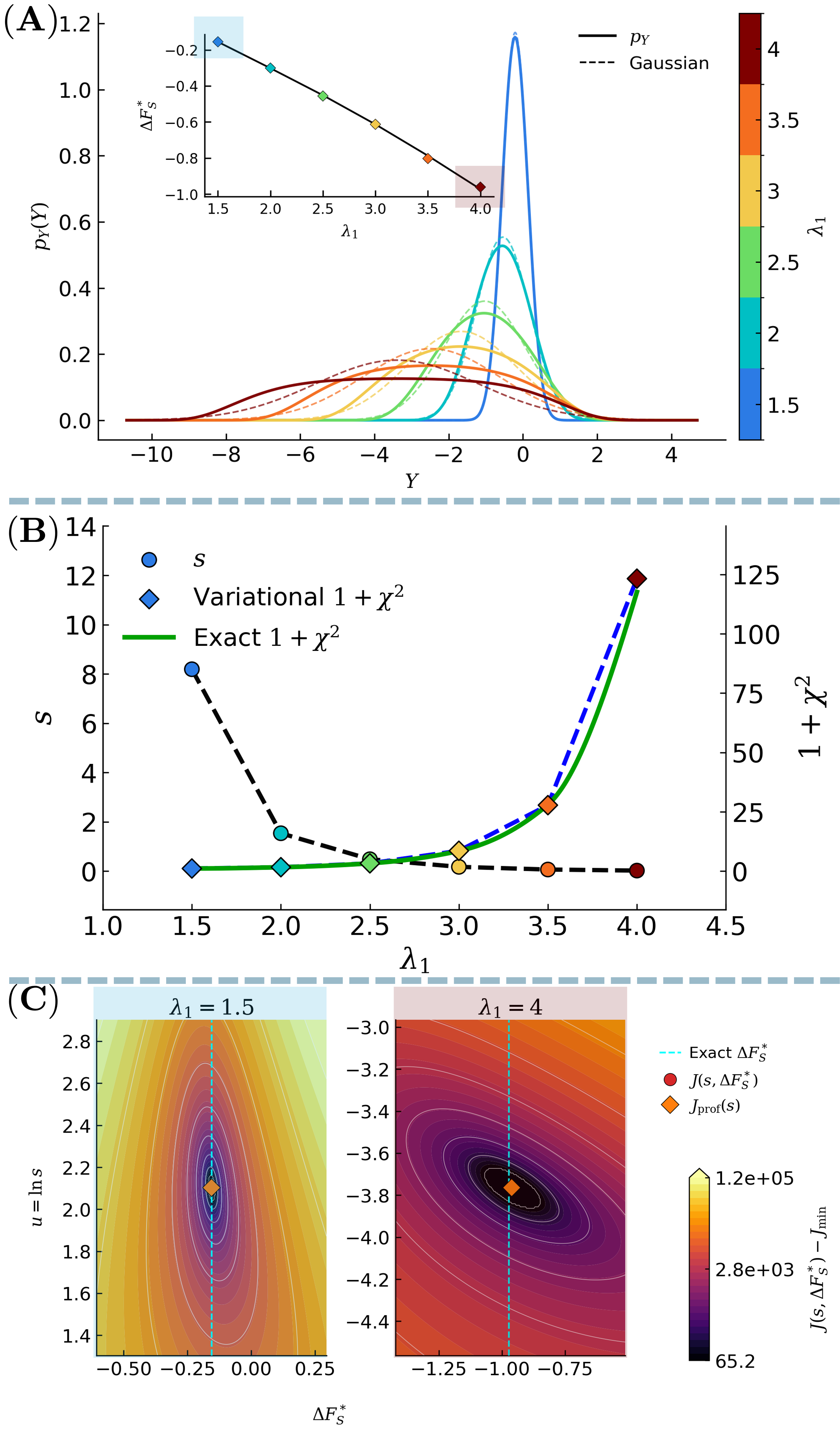}
\caption{\textbf{Variational scalar-action law.}
\textbf{(A)} Scalar-action densities $p_Y(Y)$ for different final protocols
$\lambda_1\equiv\lambda(t_{\rm eq})$.
Solid curves show the optimized maximum-entropy law
$p_Y(y;s,m)=[2K_0(s)]^{-1}\exp[-s\cosh(y-m)]$; dashed curves show Gaussian
approximations to the same sampled actions.
The inset compares the corresponding variational estimates of
$\Delta F_{\mathcal S}^{*}$ with the exact HMF reference.
The color bar labels the final protocol value $\lambda_1$.
\textbf{(B)} Joint diagnostic plot of the optimized shape parameter $s$ and
the endpoint-overlap factor $1+\chi^2$ as functions of $\lambda_1$.
Circles show $s$ on the left vertical axis; diamonds show the variational
overlap reconstructed from $(K_1(\hat s)/K_0(\hat s))^2$ on the right
vertical axis.
The green curve gives the exact endpoint overlap.
\textbf{(C)} Likelihood landscapes
$J(s,\Delta F_{\mathcal S}^{*})-J_{\min}$ for weak and strong driving,
plotted in the $(\Delta F_{\mathcal S}^{*},u)$ plane with $u=\ln s$.
Dashed cyan lines mark the exact free energies; symbols mark the optimized
values from the full two-parameter objective and from the profiled objective.}
\label{fig:variational_estimator}
\end{figure}

\begin{figure}[t!]
  \centering
  \includegraphics[width=\columnwidth]{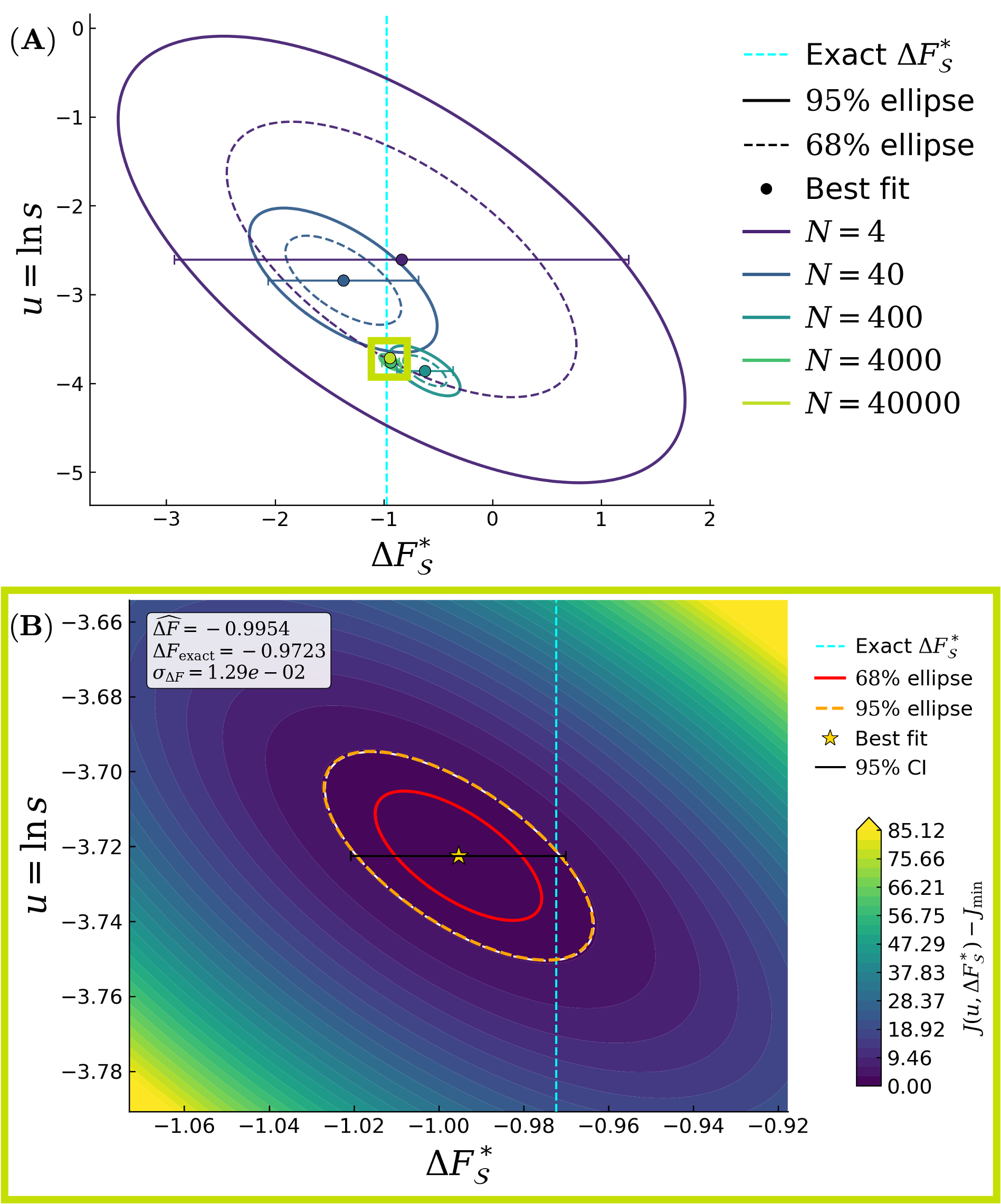}
\caption{
\textbf{Hessian uncertainty of the variational estimator.} \textbf{(A)} Nominal local confidence ellipses in the $(\Delta F_{\mathcal S}^{*},u)$ plane, with $u=\ln s$, for increasing sample size $N$.
Solid and dashed curves denote nominal $95\%$ and $68\%$ local regions.
Horizontal bars show nominal one-parameter $95\%$ local intervals for $\Delta F_{\mathcal S}^{*}$.
\textbf{(B)} Local objective landscape for $N=40000$ at $\lambda(t_{\rm eq})=4$.
The optimized value agrees with the exact HMF reference within the nominal Hessian $95\%$ local interval.}
  \label{fig:hessian_uncertainty}
\end{figure}

\noindent
\textit{Variational construction.} The endpoint and trajectory relations above are exact identities.
The maximum-entropy construction introduced below is an additional variational reconstruction used to infer a tractable scalar-action law from finite fluctuation information.
The preceding identities convert the open-system free energy problem into an inference problem for a scalar random variable $Y=\mathcal Y(\Xi)$, where $\Xi$ denotes the sampled microscopic object and $\mu(\mathrm d\Xi)$ its sampling measure.
This is the pushforward distribution of the microscopic sampling measure under the observable map $\mathcal Y$, a standard construction in probability and statistical inference \cite{cover1999elements}.
This reduction is statistical, not physical: the many-particle structure, the system--environment coupling, and the full trajectory history remain encoded in the map $\mathcal Y$; only the object to be inferred is the induced one-dimensional law $p_Y$.
In the fixed-coupling branch, $\mathcal Y$ may be evaluated from endpoint configurations ($Y=\beta\Delta\mathcal H_{\mathcal S}(X_{\mathcal S})$), from propagated initial conditions ($Y=\beta\Delta\mathcal H_{\mathcal S}(X_0)$), or from the heat--work--reference action ($Y=\beta(W_{\mathcal S}+Q_{\mathcal S}-II)$). 
With this convention, the exact identities impose
\begin{equation}
\label{eq_y_ex_pl}
\langle e^{Y}\rangle_Y
=
e^{\beta\Delta F_{\mathcal S}^{*}},
\end{equation}
and
\begin{equation}
\label{eq_y_ex_min_va}
\langle e^{-Y}\rangle_Y
=
e^{-\beta\Delta F_{\mathcal S}^{*}}
\left[
1+\chi^2(P_{t_{\rm eq}}^{\mathcal S}\parallel P_0^{\mathcal S})
\right],
\end{equation}
so that
\begin{equation}
\label{eq_ex_pro_y}
\langle e^{Y}\rangle_Y
\langle e^{-Y}\rangle_Y
=
1+\chi^2(P_{t_{\rm eq}}^{\mathcal S}\parallel P_0^{\mathcal S}).
\end{equation}
Equation~\eqref{eq_ex_pro_y} is the key diagnostic content of the scalar reduction.
The high-dimensional endpoint overlap is encoded in the one-dimensional fluctuations of $Y$. The present framework does not eliminate the statistical limitations associated with poor overlap; rather, it converts the overlap burden into an explicit inferable quantity through Eq.~\eqref{eq_ex_pro_y}.
The estimator therefore reports not only the HMF free energy, but also the statistical overlap burden supporting that estimate, a central issue in exponential free energy estimators and Bennett-type methods \cite{bennett1976efficient,shirts2008statistically, klimovich2015guidelines,pohorille2010good}.
The two exponential moments constrain $p_Y$ but do not determine its full shape.
We therefore reconstruct the least-biased compatible law by maximizing its entropy $\mathbf{S}[p_Y]=-\int \mathrm dy\,p_Y(y)\ln p_Y(y)$ under normalization and Eqs.~\eqref{eq_y_ex_pl}--\eqref{eq_y_ex_min_va}, following the maximum-entropy principle for assigning distributions from moment information \cite{jaynes1957information,shore2003axiomatic,cover1999elements,honerkamp1996stochastic}.
This variational step is independent of microscopic details once the scalar action $Y=\mathcal Y(\Xi)$ has been specified.
The Lagrange-multiplier stationarity condition gives $p_Y(y)\propto\exp[-\lambda_+e^y-\lambda_-e^{-y}]$, where $\lambda_{+/-}$ are Lagrange coefficients, the exponential-family form associated with moment-constrained maximum-entropy inference \cite{cover1999elements,wainwright2008graphical}.
With $s=2\sqrt{\lambda_+\lambda_-}$ and $m=\frac{1}{2}\ln(\lambda_-/\lambda_+)$, normalization yields
\begin{equation}
\label{eq_prob_vari_non}
p_Y(y;s,m)
=
\frac{1}{2K_0(s)}
\exp[-s\cosh(y-m)],
\end{equation}
where $K_\nu$ is the modified Bessel function of the second kind.
The normalization follows from the standard integral representation of $K_\nu$ \cite{abramowitz1948handbook}.
The derivation is given in the Appendix.
The moments of Eq.~\eqref{eq_prob_vari_non} give $\beta\Delta F_{\mathcal S}^{*} = m+\ln[{K_1(s)}/{K_0(s)}]$ and
\begin{equation}
\label{eq_chi_vari_key}
\left(\frac{K_1(s)}{K_0(s)}\right)^2
=
1+\chi^2(P_{t_{\rm eq}}^{\mathcal S}\parallel P_0^{\mathcal S}).
\end{equation}
Equation~\eqref{eq_chi_vari_key} is the central practical output of the variational construction: within the maximum-entropy scalar-action representation, the high-dimensional endpoint overlap is encoded in a single optimized Bessel parameter.
The central consequence is that the overlap burden appears as an explicit scalar quantity once the scalar-action law has been reconstructed.
Instead of remaining hidden as an implicit parameter affecting convergence, it is reported through the Bessel ratio in Eq.~\eqref{eq_chi_vari_key}.
Thus, one measured probability distribution yields the HMF free energy estimate, the endpoint-overlap diagnostic, and the scalar-action shape. Fig.~\ref{fig:variational_estimator} shows these three outputs.
For sampled values $Y_i=\mathcal Y(\Xi_i)$, we determine the two parameters of the reconstructed law by likelihood optimization, following the standard maximum-likelihood treatment of parametric statistical models \cite{fisher1922mathematical,cox2006principles,honerkamp1996stochastic}.
Written in terms of
$(s,\Delta F_{\mathcal S}^{*})$, the negative log-likelihood is 
\begin{align}
\label{eq_vari_dota}
J(s,\Delta F_{\mathcal S}^{*})
&\nonumber=
N\ln[2K_0(s)]
\\&+
s\sum_{i=1}^{N}
\cosh\!\left[
Y_i-\beta\Delta F_{\mathcal S}^{*}
+\ln\frac{K_1(s)}{K_0(s)}
\right].
\end{align}
The variational estimator is
\begin{equation}
\label{}
(\hat s,\widehat{\Delta F_{\mathcal S}^{*}})
=
\arg\min_{s>0,\;\Delta F_{\mathcal S}^{*}\in\mathbb R}
J(s,\Delta F_{\mathcal S}^{*}).
\end{equation}
The Hessian of $J$ at the optimum gives the observed-information approximation to the local covariance of the inference parameters and the standard error of $\widehat{\Delta F_{\mathcal S}^{*}}$ \cite{fisher1922mathematical,efron1978assessing}.
The profile reduction of the optimization and the Hessian uncertainty construction are given in the End Matter and in the Appendix.  Fig.~\ref{fig:hessian_uncertainty} shows the resulting local uncertainty. 
\textit{Conclusions.}
We derived exact HMF-based identities for free energy differences in strongly coupled open systems, using canonical HMF endpoint structure rather than dynamical conditions that guarantee or repair nonequilibrium work identities through microscopic reversibility, DB, FDT, LDB, or path-weight corrections.
Under the explicit asymptotic equilibration condition, these identities also admit trajectory representations, allowing endpoint free energy relations to be evaluated from trajectory-generated samples.
In the frozen-coupling regime, the standard JE form is recovered as the Liouvillian limiting case, where the HMF partition-function ratio becomes the exponential average of the nonequilibrium work \cite{jarzynski1997nonequilibrium,jarzynski2004nonequilibrium,sagawa2010generalized,talkner2020colloquium}. 
Building on these identities, we introduced a free energy estimation method based on a scalar-action law.
Its practical value is that one sampled scalar-action distribution carries three pieces of information at once: the HMF free energy difference, the endpoint-overlap burden, and the local uncertainty.
In standard FEP, BAR, MBAR, and JE estimators, the free energy estimate is obtained directly, while the overlap burden usually appears indirectly through convergence, bias, or sampling diagnostics \cite{zwanzig1954high,bennett1976efficient,shirts2008statistically,klimovich2015guidelines,pohorille2010good}.
Here, by contrast, the statistical cost of connecting the two endpoints is part of the identity itself, making the estimator overlap-resolved rather than only free energy-resolved. 
The non-Liouvillian validation confirms this distinction: the endpoint identities recover the exact HMF free energy where the standard JE work-reweighting estimator fails, and the variational scalar-action construction reproduces the same reference within its local uncertainty.
This framework therefore provides an overlap-resolved route for free energy calculations in strongly coupled open systems and suggests extensions to active matter, biological processes, complex molecular environments, and open quantum systems, where weak-coupling assumptions or idealized reversible dynamics are often unavailable \cite{gaspard2020active,gnesotto2018broken,woodside2014reconstructing,leighton2025flow,beyer2025operational}.
\vspace{-0.8cm}
\begin{acknowledgments} We gratefully acknowledge financial support by the DFG
under Germany's Excellence Strategy EXC 2089/1-390776260
(e-conversion).
\end{acknowledgments}

\bibliography{apssamp}

\newpage

\onecolumngrid
\begin{center}
\Large \bfseries End Matter
\end{center}
\twocolumngrid

\begin{figure}[htb!] 
  \centering
  \includegraphics[width=0.85\columnwidth]{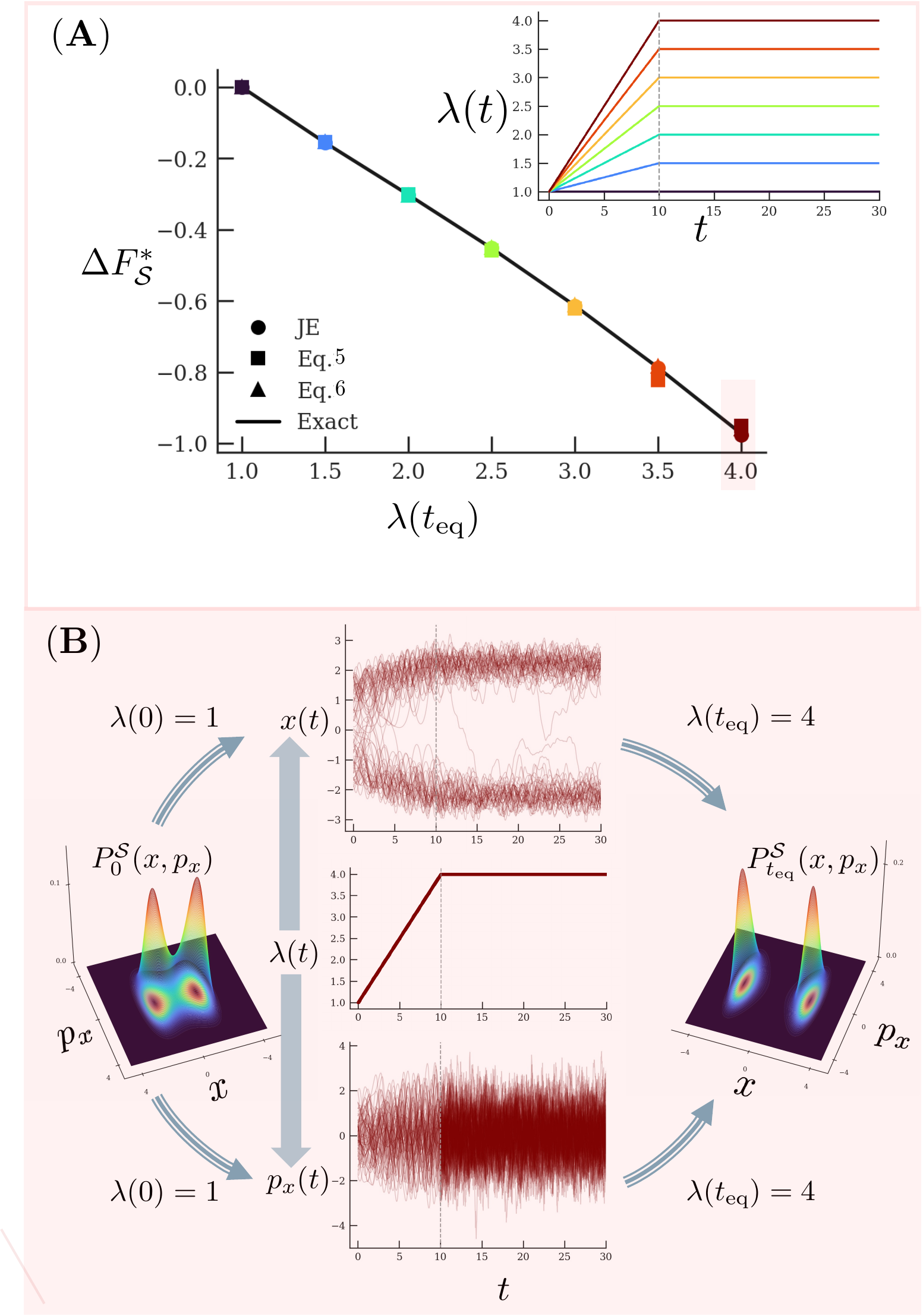}
  \caption{
  Same plotting conventions as Fig.~\ref{fig2}. During the ramp,
  $\gamma_{\rm nl}^{\mathcal S}=\gamma_{\rm nl}^{\mathcal E}=0$, so the
  ramp is Liouvillian and the JE, endpoint equalities, and exact HMF reference
  coincide. }
  \label{fig3}
\end{figure}

\begin{figure}[!]
\centering
\begin{tcolorbox}[
  width=\columnwidth,
  colback=white,
  colframe=black!60,
  arc=2mm,
  boxrule=0.85pt,
  left=3pt,
  right=3pt,
  top=4pt,
  bottom=4pt,
  boxsep=1pt
]
\centering
{\footnotesize\bfseries From work reweighting to scalar-action inference}

\vspace{0.06cm}

\begin{tcolorbox}[
  width=0.95\columnwidth,
  colback=red!7,
  colframe=red!45!black,
  arc=1.7mm,
  boxrule=0.85pt,
  left=3pt,
  right=3pt,
  top=3pt,
  bottom=2pt,
  boxsep=1pt
]
\centering
{\scriptsize\color{red!45!black}\bfseries A. JE route}

\vspace{-0.30cm}
{\scriptsize
\[
\left\langle e^{-\beta W}\right\rangle
=
e^{-\beta\Delta F_{\mathcal S}^{*}}.
\]
}
\vspace{-0.6cm}

{\scriptsize work reweighting under dynamical reversibility}
\end{tcolorbox}

\vspace{-0.1cm}

{\scriptsize\itshape If reversibility is lost, the work identity need not survive.}

\vspace{-0.1cm}

\rule{\columnwidth}{0.85pt}

\vspace{-0.1cm}

\begin{tcolorbox}[
  width=0.95\columnwidth,
  colback=blue!7,
  colframe=blue!55!black,
  arc=1.7mm,
  boxrule=0.85pt,
  left=3pt,
  right=3pt,
  top=3pt,
  bottom=2pt,
  boxsep=1pt
]
\centering
{\scriptsize\color{blue!55!black}\bfseries B. Endpoint--trajectory--overlap route}

\vspace{-0.40cm}
{\scriptsize
\[
\begin{gathered}
Y=\mathcal Y(\Xi),
\\[1pt]
\langle e^{Y}\rangle_Y
=
e^{\beta\Delta F_{\mathcal S}^{*}},
\\[1pt]
\langle e^{-Y}\rangle_Y
=
e^{-\beta\Delta F_{\mathcal S}^{*}}
\left[
1+\chi^2(P_{t_{\rm eq}}^{\mathcal S}\parallel P_0^{\mathcal S})
\right],
\\[1pt]
\langle e^{Y}\rangle_Y
\langle e^{-Y}\rangle_Y
=
1+\chi^2(P_{t_{\rm eq}}^{\mathcal S}\parallel P_0^{\mathcal S}) .
\end{gathered}
\]
}
\vspace{-0.24cm}

{\scriptsize canonical endpoints make overlap explicit}
\end{tcolorbox}

\vspace{-0.15cm}
{\small$\Downarrow$}
\vspace{-0.15cm}

\begin{tcolorbox}[
  width=0.95\columnwidth,
  colback=blue!7,
  colframe=blue!50!black,
  arc=1.7mm,
  boxrule=0.85pt,
  left=3pt,
  right=3pt,
  top=2pt,
  bottom=1pt,
  boxsep=1pt
]
\centering
{\scriptsize\color{blue!50!black}\bfseries I.Scalar-action choices}

\vspace{-0.4cm}
{\scriptsize
\[
\begin{array}{ll}
\text{endpoint:} &
Y=\beta\Delta\mathcal H_{\mathcal S}(X_{\mathcal S}),
\\[1pt]
\text{trajectory:} &
Y=\beta\Delta\mathcal H_{\mathcal S}(X_0),
\\[1pt]
\text{heat--work--reference:} &
Y=\beta(W_{\mathcal S}+Q_{\mathcal S}-II).
\end{array}
\]
}
\end{tcolorbox}

\vspace{-0.15cm}
{\small$\Downarrow$}
\vspace{-0.15cm}

\begin{tcolorbox}[
  width=0.95\columnwidth,
  colback=blue!7,
  colframe=blue!60!black,
  arc=1.7mm,
  boxrule=0.85pt,
  left=3pt,
  right=3pt,
  top=3pt,
  bottom=2pt,
  boxsep=1pt
]
\centering
{\scriptsize\color{blue!60!black}\bfseries II. Variational inference}

\vspace{-0.40cm}
{\scriptsize
\[
p_Y(y;s,m)
=
\frac{e^{-s\cosh(y-m)}}{2K_0(s)},
\]
}
\vspace{-0.10cm}
{\scriptsize
\[
K_\nu(s)
=
\int_0^\infty e^{-s\cosh t}\cosh(\nu t)\,\mathrm dt
,\]
}

\vspace{-0.22cm}
{\scriptsize
\[
J
=
N\ln[2K_0(s)]
+
s\sum_i
\cosh\!\left[
Y_i-\beta\Delta F_{\mathcal S}^{*}
+\ln\frac{K_1(s)}{K_0(s)}
\right].
\]
}
\vspace{-0.3cm}

{\scriptsize system-independent Bessel law and likelihood inference}
\end{tcolorbox}

\vspace{-0.15cm}
{\small$\Downarrow$}
\vspace{-0.15cm}

\begin{tcolorbox}[
  width=0.95\columnwidth,
  colback=blue!7,
  colframe=blue!65!black,
  arc=1.7mm,
  boxrule=0.85pt,
  left=3pt,
  right=3pt,
  top=2pt,
  bottom=1pt,
  boxsep=1pt
]
\centering
{\scriptsize\color{blue!65!black}\bfseries III.Main output: free energy estimate, overlap diagnostic, and uncertainty}

\vspace{-0.4cm}
{\scriptsize
\[
(\hat s,\widehat{\Delta F_{\mathcal S}^{*}})
=
\arg\min J,
\qquad
\widehat{1+\chi^2}_{\rm Bessel}
=
\left(\frac{K_1(\hat s)}{K_0(\hat s)}\right)^2
\]
}
\vspace{-0.8cm}

{\scriptsize
\[
\mathbf H=\nabla_{\vartheta}^{2}J|_{\hat{\vartheta}},
\qquad
\mathbf\Sigma\simeq \mathbf H^{-1},
\qquad
\sigma_{\Delta F}=\sqrt{\Sigma_{\Delta F,\Delta F}}.
\]
}
\vspace{-0.20cm}
\end{tcolorbox}

\end{tcolorbox}

\caption{
\textbf{Conceptual structure.}
The JE route relies on dynamical reversibility.
The endpoint route replaces this requirement by canonical endpoint structure and reduces the problem to a scalar action $Y$.
The endpoint identities impose the positive and negative exponential moments of $Y$, whose product gives the endpoint-overlap factor.
The variational law $p_Y$ has a constraint-universal Bessel form once the scalar action is specified.
Likelihood optimization of $J$ gives the HMF free energy estimate and the optimized Bessel parameter, while the inverse Hessian of $J$ gives the local uncertainty.}
\label{fig:conceptual_structure}
\end{figure}
\noindent
The Appendix gives the full derivations of the endpoint
and trajectory identities, the heat--work--reference decomposition, the
maximum-entropy scalar-action construction, the Bessel moment relations, the
profile reduction, the Hessian uncertainty analysis, the validation model and
numerical protocol, the Liouvillian consistency check, and the multistage
overlap-resolved comparison with BAR and MBAR. Fig.~\ref{fig:conceptual_structure} summarizes this logical structure. 
\textit{Heat--work--reference decomposition and limiting cases.} For fixed coupling, the trajectory increment in Eqs.~\eqref{eq_fixc_ratio} and \eqref{eq_fixc_momen} decomposes as $\label{eq:DeltaH_heatwork_final}
\Delta \mathcal H_{\mathcal S}(X_0)
=
W_{\mathcal S}(t_{\rm eq}|X_0)
+
Q_{\mathcal S}(t_{\rm eq}|X_0)
-
II(t_{\rm eq}|X_0)$. Here
$W_{\mathcal S}
=
\int_0^{t_{\rm eq}}\mathrm dt\,
[\partial\mathcal H_{\mathcal S}(X_{\mathcal S}(t|X_0),\lambda(t))/\partial\lambda]\dot\lambda(t)$ is the parametric work, $Q_{\mathcal S}(t_{\mathrm{eq}}|X_0) =\int_{0}^{t_{\mathrm{eq}}}{\rm d}t\,\big(\nabla_{X_{\mathcal S}}\mathcal H_{\mathcal S}\big) \cdot\dot X_{\mathcal S}(t|X_0),$ is the heat-like energy change generated by motion through phase space \cite{seifert2012stochastic}, and $II(t_{\mathrm{eq}}|X_0)=\int_0^{t_{\mathrm{eq}}}\mathrm dt\; \Big(\nabla_{X_{\mathcal S}}\mathcal H_{\mathcal S}(X_{\mathcal S},\lambda(0))\Big)\cdot\dot X_{\mathcal S}$ is the reference projection of the realized trajectory onto the initial force field.
Substitution into Eqs.~\eqref{eq_fixc_ratio} and \eqref{eq_fixc_momen} gives
\begin{align}
\label{eq_heat_w_frac}
e^{-\beta\Delta F_{\mathcal S}^{*}(\beta)}
&=
\frac{
\left\langle
e^{-\beta[W_{\mathcal S}+Q_{\mathcal S}-II]}
\right\rangle_{X_0}
}{
1+\chi^2(P^{\mathcal S}_{t_{\rm eq}}\parallel P^{\mathcal S}_0)
},
\\[3pt]
\label{eq_hea_wo_fa_fra}
e^{+\beta\Delta F_{\mathcal S}^{*}(\beta)}
&=
\left\langle
e^{+\beta[W_{\mathcal S}+Q_{\mathcal S}-II]}
\right\rangle_{X_0}.
\end{align}
The partition-function identity behind these relations is $e^{-\beta\Delta F_{\mathcal S}^{*}} = \mathcal Z_{\mathcal S}^{*}(\lambda(t_{\rm eq}),C(0),\beta)/ \mathcal Z_{\mathcal S}^{*}(\lambda(0),C(0),\beta)$, equivalently the same
composite ratio since $\mathcal Z_{\mathcal S+\mathcal E} = \mathcal Z_{\mathcal S}^{*}\mathcal Z_{\mathcal E}$ \cite{jarzynski2004nonequilibrium,talkner2020colloquium}.
This identity contains no dynamical assumption.
The standard JE is recovered when Hamiltonian composite dynamics and Liouville's theorem convert this partition-function ratio into a work exponential average \cite{jarzynski1997nonequilibrium,jarzynski2004nonequilibrium}. Fig.~\ref{fig3} confirms this Liouvillian limiting case.
Feedback extensions restore the work identity by adding stochastic mutual information under LDB \cite{sagawa2010generalized,toyabe2010experimental}.
In contrast, Eqs.~\eqref{eq_heat_w_frac} and \eqref{eq_hea_wo_fa_fra} require neither measurement feedback nor LDB.
Their structure separates the kinematic path contribution $Q_{\mathcal S}-II$ from the endpoint-overlap factor.
\textit{Profile likelihood and Hessian uncertainty.} The likelihood in Eq.~\eqref{eq_vari_dota} has a useful profile form.
For fixed $s$, the curvature in the free energy direction is
\begin{equation}
\label{}
\frac{\partial^2J}{\partial(\Delta F_{\mathcal S}^{*})^2}
=
\beta^2s
\sum_i
\cosh\!\left[
Y_i-\beta\Delta F_{\mathcal S}^{*}
+\ln\frac{K_1(s)}{K_0(s)}
\right]
>0 .
\end{equation}
The optimum in $\Delta F_{\mathcal S}^{*}$ is therefore unique for each fixed $s$. With $c(s)=\ln[K_1(s)/K_0(s)]$, stationarity gives
\begin{equation}
\label{}
\Delta F_{\mathcal S}^{*}(s)
=
\frac{1}{\beta}
\operatorname{artanh}
\frac{
\sum_i\sinh[Y_i+c(s)]
}{
\sum_i\cosh[Y_i+c(s)]
}.
\end{equation}
Thus the two-parameter inference reduces to $\hat s=\arg\min_{s>0}J(s,\Delta F_{\mathcal S}^{*}(s))$ and $\widehat{\Delta F_{\mathcal S}^{*}}=\Delta F_{\mathcal S}^{*}(\hat s)$.
For the uncertainty estimate, we set $u=\ln s$ and $\boldsymbol\vartheta=(u,\Delta F_{\mathcal S}^{*})^{\rm T}$.
The local quadratic expansion $J(\boldsymbol\vartheta)\simeq J(\hat{\boldsymbol\vartheta}) + \frac{1}{2} (\boldsymbol\vartheta-\hat{\boldsymbol\vartheta})^{\rm T} \mathbf H (\boldsymbol\vartheta-\hat{\boldsymbol\vartheta})$ defines $\mathbf H=\nabla_{\boldsymbol\vartheta}^2J|_{\hat{\boldsymbol\vartheta}}$.
The covariance estimate is $\mathbf\Sigma\simeq\mathbf H^{-1}$, and the local standard error is $\sigma_{\Delta F}=\sqrt{\Sigma_{\Delta F,\Delta F}}$.
This quantifies how uncertainty in the free energy couples to uncertainty in the endpoint-overlap parameter \cite{fisher1922mathematical,cox2006principles,efron1978assessing}.
\textit{Bessel structure of the scalar-action law.} The Bessel functions enter only through normalization and moments of the maximum-entropy density.
The stationary density $p_Y(y)\propto \exp[-\lambda_+e^y-\lambda_-e^{-y}]$ becomes $p_Y(y)\propto\exp[-s\cosh(y-m)]$ after setting $s=2\sqrt{\lambda_+\lambda_-}$ and $m=\frac{1}{2}\ln(\lambda_-/\lambda_+)$.
Since $\int_{-\infty}^{\infty}\mathrm dy\,e^{-s\cosh(y-m)}=2K_0(s)$, $p_Y(y;s,m)=[2K_0(s)]^{-1}e^{-s\cosh(y-m)}$.
Its moments satisfy $\langle e^{\pm Y}\rangle_Y=e^{\pm m}K_1(s)/K_0(s)$.
Multiplying both moments cancels $m$ and gives $\langle e^{Y}\rangle_Y\langle e^{-Y}\rangle_Y=[K_1(s)/K_0(s)]^2$.
Comparison with Eq.~\eqref{eq_ex_pro_y} gives the overlap--Bessel relation $1+\chi^2(P_{t_{\rm eq}}^{\mathcal S}\parallel P_0^{\mathcal S}) =[K_1(s)/K_0(s)]^2$ within the maximum-entropy scalar-action representation.
The Bessel mapping is independent of the microscopic system and environment; their role is to generate the sampled scalar actions $Y_i$ from which the optimized parameter $\hat s$ is obtained \cite{abramowitz1948handbook}.
\textit{Numerical validation model and protocol.} We validate the fixed-coupling identities on a composite system consisting of a double-well system bilinearly coupled to a harmonic environment \cite{whitelam2025improving,hanggi1990reaction,li2021equilibrium, saito1976relaxation,sun2003equilibrium}.
The system Hamiltonian is $\mathcal H_{\mathcal S}(x,p_x;\lambda)=p_x^2/(2m)+U_{\mathcal S}(x;\lambda)$ with $U_{\mathcal S}(x;\lambda)=\frac14(x^2-\lambda)^2$; the environment is $\mathcal H_{\mathcal E}(y,p_y)=p_y^2/(2m)+\frac12\omega^2y^2$, and the interaction is $\mathcal V_{\mathcal S\mathcal E}(x,y;C)=Cxy$.
Here $y$ and $p_y$ are the environment coordinate and momentum, $\omega$ is the environment frequency, and $m$ is the mass.
The composite is initialized canonically at $(\lambda(0),C(0))$.
We then use a two-stage protocol. 
During the ramp, $C(t)\equiv C(0)$ and $\lambda(t)$ changes linearly from $\lambda(0)$ to $\lambda(t_\lambda)$, where $t_\lambda$ is the end of the ramp.
The composite evolves under the non-Liouvillian deterministic equations $\dot x=p_x/m$, $\dot p_x=-\partial_xU_{\mathcal S}(x;\lambda(t))-Cy-\gamma_{\rm nl}^{\mathcal S}p_x/m$, $\dot y=p_y/m$, and $\dot p_y=-\omega^2y-Cx-\gamma_{\rm nl}^{\mathcal E}p_y/m$, where $\gamma_{\rm nl}^{\mathcal S}$ and $\gamma_{\rm nl}^{\mathcal E}$ are non-Liouvillian drag coefficients for the system and environment.
After the ramp, $\lambda$ is held fixed and the composite relaxes under underdamped Langevin dynamics obeying the FDT \cite{hanggi1990reaction,zwanzig2001nonequilibrium}.
This relaxation prepares the final canonical endpoint and satisfies the asymptotic equilibration condition used in the main text.
The exact reference curve in Fig.~\ref{fig2} is obtained from the HMF partition-function ratio. The JE estimator uses the ramp work $W_\lambda=\int_0^{t_\lambda} \partial_\lambda U_{\mathcal S}(x_t;\lambda_t)\dot\lambda\,\mathrm dt$ and $e^{-\beta\Delta F_{\mathcal S}^{*}}=\langle e^{-\beta W_\lambda}\rangle$.
Because the ramp is non-Liouvillian, the standard JE estimator without a phase-space-compression correction deviates from the exact HMF free energy, whereas Eqs.~\eqref{eq_fixc_ratio} and \eqref{eq_fixc_momen} recover the equilibrium reference.
The same sampled trajectories define the scalar actions $Y_i$ used in the variational construction.
The optimized density in Eq.~\eqref{eq_prob_vari_non} captures the sampled non-Gaussian distribution laws, and the optimized Bessel parameter provides the corresponding endpoint-overlap diagnostic through Eq.~\eqref{eq_chi_vari_key}.
For the representative protocol $\lambda(0)=1\to\lambda(t_{\rm eq})=4$, the optimized value $\widehat{\Delta F}_{\mathcal S}^{*}=-0.9954$ agrees with the exact reference $\Delta F_{\mathcal S,{\rm exact}}^{*}=-0.9723$ within the Hessian $95\%$ interval $[-1.0207,-0.9700]$, with $\sigma_{\Delta F}=1.29\times10^{-2}$.

\onecolumngrid
\appendix
\section{Scope and notation}
\label{sec:SM_scope}

\noindent
This Appendix gives the derivations, numerical definitions, and
additional comparisons supporting the main text.
The first part fixes the notation and summarizes the logical structure of the
construction.
The second part derives the endpoint HMF fluctuation identities and their
trajectory representation.
The third part specializes these identities to frozen coupling, where the HMF
increment becomes a bare-system increment and admits a heat--work--reference
decomposition.
The fourth part develops the scalar-action formulation, including the
maximum-entropy reconstruction, Bessel moment structure, likelihood estimator,
profile reduction, and Hessian uncertainty estimate.
The final part gives the validation model, numerical protocol, Liouvillian
consistency check, multistage overlap-resolved reconstruction, and comparison
with FEP, BAR, and MBAR. 
The composite phase point is denoted by
\begin{equation}
X=(X_{\mathcal S},X_{\mathcal E}),
\end{equation}
where $X_{\mathcal S}$ and $X_{\mathcal E}$ collect the system and environment degrees of freedom.
The control parameter $\lambda(t)$ drives the bare system Hamiltonian, while $C(t)$ controls the system--environment interaction.
The total Hamiltonian is
\begin{align}
\label{eq_totham}
\mathcal H_{\mathcal S+\mathcal E}(X,\lambda,C)
&=
\mathcal H_{\mathcal S}(X_{\mathcal S},\lambda)
+
\mathcal H_{\mathcal E}(X_{\mathcal E})
+
\mathcal V_{\mathcal S\mathcal E}(X_{\mathcal S},X_{\mathcal E};C).
\end{align}
For fixed $(\lambda,C,\beta)$, the composite canonical distribution is
\begin{equation}
\label{eq_compo_pro}
P(X;\lambda,C,\beta)
=
\frac{
e^{-\beta \mathcal H_{\mathcal S+\mathcal E}(X,\lambda,C)}
}{
\mathcal Z_{\mathcal S+\mathcal E}(\lambda,C,\beta)
},
\end{equation}
where
\begin{equation}
\label{}
\mathcal Z_{\mathcal S+\mathcal E}(\lambda,C,\beta)
=
\int \mathrm dX\,
e^{-\beta \mathcal H_{\mathcal S+\mathcal E}(X,\lambda,C)}.
\end{equation}
The open system is described by the marginal distribution of $\mathcal S$.
At strong coupling, this marginal is not governed only by the bare Hamiltonian $\mathcal H_{\mathcal S}$. 
It is governed by the Hamiltonian of mean force (HMF), which incorporates the equilibrium influence of the environment \cite{jarzynski2004nonequilibrium,talkner2020colloquium}.
The HMF marginal is
\begin{equation}
\label{eq_hmf_pro}
P^{\mathcal S}(X_{\mathcal S};\lambda,C,\beta)
=
\frac{
e^{-\beta\mathcal H_\beta^*(X_{\mathcal S},\lambda,C)}
}{
\mathcal Z_{\mathcal S}^*(\lambda,C,\beta)
},
\end{equation}
with
\begin{align}
\label{eq_hmf_def}
\mathcal H_\beta^*(X_{\mathcal S},\lambda,C)
&=
\mathcal H_{\mathcal S}(X_{\mathcal S},\lambda)
-\frac{1}{\beta}
\ln
\int \mathrm dX_{\mathcal E}\,
\frac{
e^{-\beta[
\mathcal H_{\mathcal E}(X_{\mathcal E})
+
\mathcal V_{\mathcal S\mathcal E}(X_{\mathcal S},X_{\mathcal E};C)
]}
}{
\mathcal Z_{\mathcal E}
},
\end{align}
and
\begin{equation}
\label{}
\mathcal Z_{\mathcal S}^*(\lambda,C,\beta)
=
\int \mathrm dX_{\mathcal S}\,
e^{-\beta\mathcal H_\beta^*(X_{\mathcal S},\lambda,C)}.
\end{equation}
Here
\begin{equation}
\label{}
\mathcal Z_{\mathcal E}
=
\int \mathrm dX_{\mathcal E}\,
e^{-\beta\mathcal H_{\mathcal E}(X_{\mathcal E})}
\end{equation}
is the bare environment partition function.
The corresponding HMF free energy is
\begin{equation}
\label{}
F_{\mathcal S}^*(\lambda,C,\beta)
=
-\frac{1}{\beta}\ln \mathcal Z_{\mathcal S}^*(\lambda,C,\beta).
\end{equation}

\noindent
The initial and final endpoint marginals are
\begin{align}
P_0^{\mathcal S}(X_{\mathcal S},\beta)
&=
P^{\mathcal S}(X_{\mathcal S};\lambda(0),C(0),\beta),
\\
P_{t_{\rm eq}}^{\mathcal S}(X_{\mathcal S},\beta)
&=
P^{\mathcal S}(X_{\mathcal S};\lambda(t_{\rm eq}),C(t_{\rm eq}),\beta).
\end{align}
The HMF free energy difference is
\begin{align}
\label{eq_free_def}
\Delta F_{\mathcal S}^*(\beta)
&=
F_{\mathcal S}^*(\lambda(t_{\rm eq}),C(t_{\rm eq}),\beta)-
F_{\mathcal S}^*(\lambda(0),C(0),\beta).
\end{align}
The endpoint HMF increment is
\begin{align}
\label{eq_h_star_def}
\Delta \mathcal H_\beta^*(X_{\mathcal S})
&=
\mathcal H_\beta^*(X_{\mathcal S},\lambda(t_{\rm eq}),C(t_{\rm eq}))-
\mathcal H_\beta^*(X_{\mathcal S},\lambda(0),C(0)).
\end{align}
Averages over the final system marginal are denoted by
\begin{equation}
\label{eq_aveg_s}
\langle A\rangle_{\mathcal S}
=
\int \mathrm dX_{\mathcal S}\,
A(X_{\mathcal S})P_{t_{\rm eq}}^{\mathcal S}(X_{\mathcal S},\beta).
\end{equation}
The endpoint-overlap factor is
\begin{equation}
\label{eq_chi_dive}
1+\chi^2(P_{t_{\rm eq}}^{\mathcal S}\parallel P_0^{\mathcal S})
=
\int \mathrm dX_{\mathcal S}\,
\frac{
\left[P_{t_{\rm eq}}^{\mathcal S}(X_{\mathcal S},\beta)\right]^2
}{
P_0^{\mathcal S}(X_{\mathcal S},\beta)
}.
\end{equation}
This quantity equals one only when the two endpoint marginals coincide.
Larger values indicate stronger mismatch between the final and initial system ensembles.
In the main text, this factor appears as the exact statistical correction that accompanies the negative exponential average.

\section{Endpoint fluctuation identities}
\label{sec:SM_endpoint}

\begin{theorem}[Endpoint HMF identities]
\label{thm:SM_endpoint}
For two canonical endpoint marginals at the same inverse temperature $\beta$,
\begin{align}
\label{eq_end_minu}
e^{-\beta \Delta F_{\mathcal S}^*}
&=
\frac{
\left\langle
e^{-\beta\Delta\mathcal H_\beta^*(X_{\mathcal S})}
\right\rangle_{\mathcal S}
}{
1+\chi^2(P_{t_{\rm eq}}^{\mathcal S}\parallel P_0^{\mathcal S})
},
\\
\label{eq_end_plus}
e^{+\beta \Delta F_{\mathcal S}^*}
&=
\left\langle
e^{+\beta\Delta\mathcal H_\beta^*(X_{\mathcal S})}
\right\rangle_{\mathcal S}.
\end{align}
\end{theorem}

\begin{proof}
From Eq.~\eqref{eq_hmf_pro}, the endpoint marginals obey
\begin{align}
P_0^{\mathcal S}(X_{\mathcal S},\beta)
&=
\frac{
e^{-\beta\mathcal H_\beta^*(X_{\mathcal S},\lambda(0),C(0))}
}{
\mathcal Z_{\mathcal S}^*(\lambda(0),C(0),\beta)
},
\\
P_{t_{\rm eq}}^{\mathcal S}(X_{\mathcal S},\beta)
&=
\frac{
e^{-\beta\mathcal H_\beta^*(X_{\mathcal S},\lambda(t_{\rm eq}),C(t_{\rm eq}))}
}{
\mathcal Z_{\mathcal S}^*(\lambda(t_{\rm eq}),C(t_{\rm eq}),\beta)
}.
\end{align}
Using Eq.~\eqref{eq_h_star_def}, one obtains
\begin{equation}
\label{eq_key_min}
e^{-\beta\Delta\mathcal H_\beta^*(X_{\mathcal S})}
=
\frac{
P_{t_{\rm eq}}^{\mathcal S}(X_{\mathcal S},\beta)
\mathcal Z_{\mathcal S}^*(\lambda(t_{\rm eq}),C(t_{\rm eq}),\beta)
}{
P_0^{\mathcal S}(X_{\mathcal S},\beta)
\mathcal Z_{\mathcal S}^*(\lambda(0),C(0),\beta)
}.
\end{equation}
Averaging Eq.~\eqref{eq_key_min} over $P_{t_{\rm eq}}^{\mathcal S}$ gives
\begin{align}
\left\langle e^{-\beta\Delta\mathcal H_\beta^*}\right\rangle_{\mathcal S}
&=
\frac{
\mathcal Z_{\mathcal S}^*(\lambda(t_{\rm eq}),C(t_{\rm eq}),\beta)
}{
\mathcal Z_{\mathcal S}^*(\lambda(0),C(0),\beta)
}
\int \mathrm dX_{\mathcal S}\,
\frac{
\left[P_{t_{\rm eq}}^{\mathcal S}(X_{\mathcal S},\beta)\right]^2
}{
P_0^{\mathcal S}(X_{\mathcal S},\beta)
}.
\end{align}
Equations~\eqref{eq_free_def} and \eqref{eq_chi_dive} then yield
\begin{equation}
\left\langle e^{-\beta\Delta\mathcal H_\beta^*}\right\rangle_{\mathcal S}
=
e^{-\beta\Delta F_{\mathcal S}^*}
\left[
1+\chi^2(P_{t_{\rm eq}}^{\mathcal S}\parallel P_0^{\mathcal S})
\right],
\end{equation}
which proves Eq.~\eqref{eq_end_minu}. 
For the positive exponential, the same definitions give
\begin{equation}
\label{eq:SM_key_plus}
e^{+\beta\Delta\mathcal H_\beta^*(X_{\mathcal S})}
=
\frac{
P_0^{\mathcal S}(X_{\mathcal S},\beta)
\mathcal Z_{\mathcal S}^*(\lambda(0),C(0),\beta)
}{
P_{t_{\rm eq}}^{\mathcal S}(X_{\mathcal S},\beta)
\mathcal Z_{\mathcal S}^*(\lambda(t_{\rm eq}),C(t_{\rm eq}),\beta)
}.
\end{equation}
Averaging Eq.~\eqref{eq:SM_key_plus} over $P_{t_{\rm eq}}^{\mathcal S}$ cancels the final marginal and gives
\begin{align}
\left\langle e^{+\beta\Delta\mathcal H_\beta^*}\right\rangle_{\mathcal S}
&=
\frac{
\mathcal Z_{\mathcal S}^*(\lambda(0),C(0),\beta)
}{
\mathcal Z_{\mathcal S}^*(\lambda(t_{\rm eq}),C(t_{\rm eq}),\beta)
}=
e^{+\beta\Delta F_{\mathcal S}^*}.
\end{align}
This proves Eq.~\eqref{eq_end_plus}.
\end{proof}

\noindent
Multiplying Eqs.~\eqref{eq_end_minu} and \eqref{eq_end_plus} yields the exact product relation
\begin{equation}
\label{}
\left\langle e^{+\beta\Delta\mathcal H_\beta^*}\right\rangle_{\mathcal S}
\left\langle e^{-\beta\Delta\mathcal H_\beta^*}\right\rangle_{\mathcal S}
=
1+\chi^2(P_{t_{\rm eq}}^{\mathcal S}\parallel P_0^{\mathcal S}).
\end{equation}
This identity is the source of the scalar-action overlap relation used in the main text.

\section{Trajectory representation}
\label{sec:SM_trajectory}

\noindent
Let $\mathcal T_t:\Gamma_0\to\Gamma_t$ denote the trajectory map of the composite system.
No particular generator is assumed.
The map may represent Hamiltonian, deterministic dissipative, stochastic, or more general dynamics at the level of the realized trajectory ensemble.
For notational compactness, stochastic realizations are not written as separate arguments of $\mathcal T_t$; they are understood to be included in the generated trajectory sample.
An explicit path-space formulation with the noise history as an additional variable is a natural extension, but it is not needed for the endpoint identities derived below.
The evolved composite density is written as the pushforward of the initial canonical density,
\begin{equation}
\label{}
P_t(X,\beta)
=
\int_{\Gamma_0}\mathrm dX_0\,
P_0(X_0,\beta)\,
\delta\!\left(X-\mathcal T_t(X_0)\right).
\end{equation}
The Dirac delta expresses the fact that each sampled initial state contributes probability mass at its evolved image in the realized trajectory ensemble.
Integrating over the environmental coordinates gives the system marginal,
\begin{equation}
\label{eq_push_fo}
P_t^{\mathcal S}(X_{\mathcal S},\beta)
=
\int_{\Gamma_0}\mathrm dX_0\,
P_0(X_0,\beta)\,
\delta\!\left(X_{\mathcal S}-\mathcal T_t^{\mathcal S}(X_0)\right).
\end{equation}
Indeed,
\begin{align}
P_t^{\mathcal S}(X_{\mathcal S},\beta)
&=
\int \mathrm dX_{\mathcal E}\,
P_t(X_{\mathcal S},X_{\mathcal E},\beta)
\nonumber\\
&=
\int\mathrm dX_{\mathcal E}
\int_{\Gamma_0}\mathrm dX_0\,
P_0(X_0,\beta)
\delta\!\left(X_{\mathcal S}-\mathcal T_t^{\mathcal S}(X_0)\right)
\delta\!\left(X_{\mathcal E}-\mathcal T_t^{\mathcal E}(X_0)\right)
\nonumber\\
&=
\int_{\Gamma_0}\mathrm dX_0\,
P_0(X_0,\beta)
\delta\!\left(X_{\mathcal S}-\mathcal T_t^{\mathcal S}(X_0)\right).
\end{align}

\noindent
We impose asymptotic equilibration as an endpoint condition:
\begin{equation}
\label{eq_asymp}
\lim_{t\to t_{\rm eq}}
P_t^{\mathcal S}(X_{\mathcal S},\beta)
=
P_{t_{\rm eq}}^{\mathcal S}(X_{\mathcal S},\beta).
\end{equation}
This condition states that after the driving has stopped and the final controls are held fixed, the system marginal relaxes to the final canonical HMF marginal.
It is not a reversibility, detailed-balance, or Liouville condition on the driven part of the process.
This distinction is essential.
The endpoint free energy is an equilibrium quantity, whereas the path used to reach the endpoint may be non-Hamiltonian or phase-space compressing.

\begin{theorem}[Trajectory form]
\label{thm:SM_traj}
Under Eq.~\eqref{eq_asymp},
\begin{align}
\label{eq_traj_mins}
e^{-\beta\Delta F_{\mathcal S}^*}
&=
\frac{
\left\langle
e^{-\beta\Delta\mathcal H_\beta^*
(\mathcal T_{t_{\rm eq}}^{\mathcal S}(X_0))}
\right\rangle_{X_0}
}{
1+\chi^2(P_{t_{\rm eq}}^{\mathcal S}\parallel P_0^{\mathcal S})
},
\\
\label{eq_traj_plus}
e^{+\beta\Delta F_{\mathcal S}^*}
&=
\left\langle
e^{+\beta\Delta\mathcal H_\beta^*
(\mathcal T_{t_{\rm eq}}^{\mathcal S}(X_0))}
\right\rangle_{X_0},
\end{align}
where
\begin{equation}
\langle A\rangle_{X_0}
=
\int_{\Gamma_0}\mathrm dX_0\,A(X_0)P_0(X_0,\beta).
\end{equation}
For stochastic simulations, this compact average refers to the realized trajectory ensemble generated from the initial distribution.
The random numbers or noise histories used to generate the trajectories are part of the sampling procedure, although they are not written as independent variables.
\end{theorem}

\begin{proof}
For the negative exponential, use Eqs.~\eqref{eq_aveg_s}, \eqref{eq_asymp}, and \eqref{eq_push_fo}:
\begin{align}
\left\langle
e^{-\beta\Delta\mathcal H_\beta^*(X_{\mathcal S})}
\right\rangle_{\mathcal S}
&\nonumber=
\int \mathrm dX_{\mathcal S}\,
e^{-\beta\Delta\mathcal H_\beta^*(X_{\mathcal S})}
P_{t_{\rm eq}}^{\mathcal S}(X_{\mathcal S},\beta)
\nonumber\\
&=
\lim_{t\to t_{\rm eq}}
\int \mathrm dX_{\mathcal S}\,
e^{-\beta\Delta\mathcal H_\beta^*(X_{\mathcal S})}
P_t^{\mathcal S}(X_{\mathcal S},\beta)
\nonumber\\
&=
\lim_{t\to t_{\rm eq}}
\int \mathrm dX_{\mathcal S}
\int_{\Gamma_0}\mathrm dX_0\,
P_0(X_0,\beta)
e^{-\beta\Delta\mathcal H_\beta^*(X_{\mathcal S})}
\delta\!\left(X_{\mathcal S}-\mathcal T_t^{\mathcal S}(X_0)\right)
\nonumber\\
&=
\int_{\Gamma_0}\mathrm dX_0\,
P_0(X_0,\beta)
e^{-\beta\Delta\mathcal H_\beta^*
(\mathcal T_{t_{\rm eq}}^{\mathcal S}(X_0))}
\nonumber\\
&=
\left\langle
e^{-\beta\Delta\mathcal H_\beta^*
(\mathcal T_{t_{\rm eq}}^{\mathcal S}(X_0))}
\right\rangle_{X_0}.
\end{align}
Substitution into Eq.~\eqref{eq_end_minu} proves Eq.~\eqref{eq_traj_mins}.
The same steps applied to the positive exponential prove Eq.~\eqref{eq_traj_plus}.
\end{proof}

\section{Frozen-coupling reduction}
\label{sec:SM_fixedC}

\begin{figure}[h]
    \centering
\includegraphics[width=\columnwidth]{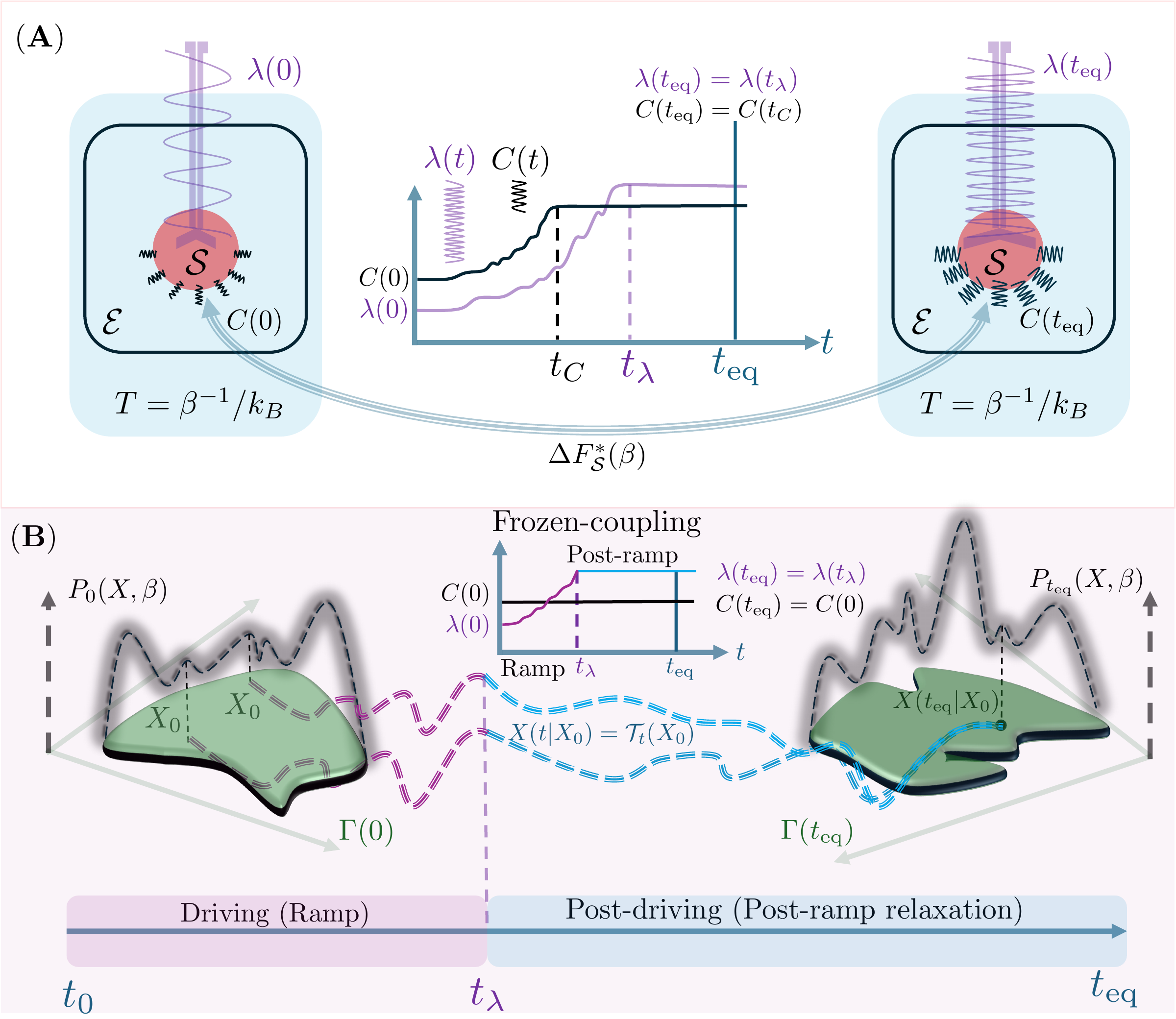}
\caption{
\textbf{Endpoint and trajectory setting.}
\textbf{(A)} Composite $\mathcal S+\mathcal E$ at fixed bath temperature
$T=(k_B\beta)^{-1}$.  The driving control $\lambda(t)$ and coupling control
$C(t)$ become constant at $t_\lambda$ and $t_C$, after which the composite
relaxes to equilibrium at $t_{\rm eq}$.  The HMF free energy difference
$\Delta F_{\mathcal S}^*(\beta)$ is defined between the canonical endpoint
marginals associated with
$(\lambda(0),C(0))$ and $(\lambda(t_{\rm eq}),C(t_{\rm eq}))$.
\textbf{(B)} An initial microstate
$X_0\sim P_0(X,\beta)$ evolves under the kinematic map
$X(t|X_0)=\mathcal T_t(X_0)$ through the driven and relaxation stages.  The
asymptotic density at $t_{\rm eq}$ defines the final canonical endpoint.  The
validation in this work uses the frozen-coupling branch,
$C(t)\equiv C(0)$.
}
    \label{fig1}
\end{figure}

\noindent
The main text focuses on the frozen-coupling branch,
\begin{equation}
\label{}
C(t)\equiv C(0),
\end{equation}
while $\lambda(t)$ is driven.
In this regime, the environmental contribution inside the logarithm in Eq.~\eqref{eq_hmf_def} is independent of $\lambda$.
Therefore, it cancels in the HMF difference taken at the same final system coordinate.
We obtain
\begin{align}
\label{eq:SM_DH_fixedC}
\Delta\mathcal H_\beta^*
(\mathcal T_{t_{\rm eq}}^{\mathcal S}(X_0))
&\nonumber=
\mathcal H_\beta^*
(\mathcal T_{t_{\rm eq}}^{\mathcal S}(X_0),\lambda(t_{\rm eq}),C(0))
-
\mathcal H_\beta^*
(\mathcal T_{t_{\rm eq}}^{\mathcal S}(X_0),\lambda(0),C(0))
\nonumber\\
&=
\mathcal H_{\mathcal S}
(\mathcal T_{t_{\rm eq}}^{\mathcal S}(X_0),\lambda(t_{\rm eq}))
-
\mathcal H_{\mathcal S}
(\mathcal T_{t_{\rm eq}}^{\mathcal S}(X_0),\lambda(0)).
\end{align}
We denote this bare-system increment by
\begin{align}
\label{eq_bare_dh}
\Delta\mathcal H_{\mathcal S}(X_0)
&=
\mathcal H_{\mathcal S}
(\mathcal T_{t_{\rm eq}}^{\mathcal S}(X_0),\lambda(t_{\rm eq}))
-
\mathcal H_{\mathcal S}
(\mathcal T_{t_{\rm eq}}^{\mathcal S}(X_0),\lambda(0)).
\end{align}
Equations~\eqref{eq_traj_mins} and \eqref{eq_traj_plus} reduce to
\begin{align}
\label{eq_fix_cmin}
e^{-\beta\Delta F_{\mathcal S}^*}
&=
\frac{
\left\langle
e^{-\beta\Delta\mathcal H_{\mathcal S}(X_0)}
\right\rangle_{X_0}
}{
1+\chi^2(P_{t_{\rm eq}}^{\mathcal S}\parallel P_0^{\mathcal S})
},
\\
\label{eq_fix_c_plu}
e^{+\beta\Delta F_{\mathcal S}^*}
&=
\left\langle
e^{+\beta\Delta\mathcal H_{\mathcal S}(X_0)}
\right\rangle_{X_0}.
\end{align}
These are the fixed-coupling identities used in the main text. Fig.~\ref{fig1} summarizes the endpoint and trajectory setting.
\section{Heat--work--reference decomposition}
\label{sec:SM_heat_work}

\noindent
We now decompose the bare-system increment in Eq.~\eqref{eq_bare_dh}.
The decomposition is algebraic, but it is useful because it separates the scalar action into three trajectory-level terms: a work-like contribution associated with the explicit protocol change, a heat-like contribution associated with motion through phase space at the instantaneous protocol, and a reference-projection contribution evaluated with the initial protocol.
The term "heat-like" is used here in a kinematic sense.
It denotes the part of the bare-system energy change induced by motion through phase space at fixed instantaneous protocol.
It coincides with thermodynamic heat only under the usual stochastic-energetics conditions and with the appropriate stochastic integral convention.
Likewise, the reference-projection contribution plays the algebraic role of a feedback-like subtraction, but it does not imply measurement, information acquisition, or feedback control. 
For each trajectory, insert and subtract the initial bare-system energy at the initial protocol,
\begin{equation}
\mathcal H_{\mathcal S}(X_{\mathcal S}(0|X_0),\lambda(0)).
\end{equation}
Using $X_{\mathcal S}(t|X_0)=\mathcal T_t^{\mathcal S}(X_0)$, Eq.~\eqref{eq_bare_dh} becomes
\begin{align}
\Delta\mathcal H_{\mathcal S}(X_0)
&=
\Big[
\mathcal H_{\mathcal S}(X_{\mathcal S}(t_{\rm eq}|X_0),\lambda(t_{\rm eq}))
-
\mathcal H_{\mathcal S}(X_{\mathcal S}(0|X_0),\lambda(0))
\Big]
\nonumber\\
&-
\Big[
\mathcal H_{\mathcal S}(X_{\mathcal S}(t_{\rm eq}|X_0),\lambda(0))
-
\mathcal H_{\mathcal S}(X_{\mathcal S}(0|X_0),\lambda(0))
\Big].
\label{eq_split_dh}
\end{align}
The first bracket is the total change of the bare-system Hamiltonian along the actual protocol.  It can be split by the chain rule:
\begin{align}
\label{eq_dh_chainr}
\frac{\mathrm d}{\mathrm dt}
\mathcal H_{\mathcal S}(X_{\mathcal S}(t|X_0),\lambda(t))
&=
\nabla_{X_{\mathcal S}}\mathcal H_{\mathcal S}
(X_{\mathcal S}(t|X_0),\lambda(t))
\cdot
\dot X_{\mathcal S}(t|X_0)
+
\frac{\partial\mathcal H_{\mathcal S}}
{\partial\lambda}
(X_{\mathcal S}(t|X_0),\lambda(t))
\dot\lambda(t).
\end{align}
This defines the work-like contribution
\begin{equation}
\label{}
W_{\mathcal S}(t_{\rm eq}|X_0)
=
\int_0^{t_{\rm eq}}\mathrm dt\,
\frac{\partial\mathcal H_{\mathcal S}}
{\partial\lambda}
(X_{\mathcal S}(t|X_0),\lambda(t))
\dot\lambda(t),
\end{equation}
and the heat-like contribution
\begin{equation}
Q_{\mathcal S}
=
\int_0^{t_{\rm eq}}\mathrm dt\,
\nabla_{X_{\mathcal S}}\mathcal H_{\mathcal S}
(X_{\mathcal S}(t|X_0),\lambda(t))
\cdot
\dot X_{\mathcal S}(t|X_0).
\end{equation}
Thus $W_{\mathcal S}$ measures the explicit energy change caused by the protocol variation $\dot\lambda(t)$, whereas $Q_{\mathcal S}$ measures the kinematic energy change caused by motion through phase space at the instantaneous protocol.
Integrating Eq.~\eqref{eq_dh_chainr} gives
\begin{align}
&\mathcal H_{\mathcal S}(X_{\mathcal S}(t_{\rm eq}|X_0),\lambda(t_{\rm eq}))
-
\mathcal H_{\mathcal S}(X_{\mathcal S}(0|X_0),\lambda(0))
=
W_{\mathcal S}(t_{\rm eq}|X_0)
+
Q_{\mathcal S}(t_{\rm eq}|X_0).
\end{align}
The second bracket in Eq.~\eqref{eq_split_dh} is evaluated with the protocol frozen at $\lambda(0)$, but along the same realized trajectory.
It defines the reference-projection functional
\begin{equation}
\label{eq:SM_II_def}
II(t_{\mathrm{eq}}|X_0)=\int_0^{t_{\mathrm{eq}}}\mathrm dt\;
\Big(\nabla_{X_{\mathcal S}}\mathcal H_{\mathcal S}(X_{\mathcal S},\lambda(0))\Big)\cdot\dot X_{\mathcal S}.
\end{equation} 

\noindent
This term projects the realized
trajectory onto the force field of the initial protocol.
It is therefore not
the physical heat exchanged along the actual protocol.
It becomes heat-like
only if the actual protocol is also fixed at $\lambda(0)$.
Its role is to
subtract the reference energy change that would be assigned to the same
trajectory under the initial protocol. 
Therefore,
\begin{equation}
\label{eq:SM_DH_WQI}
\Delta\mathcal H_{\mathcal S}(X_0)
=
W_{\mathcal S}(t_{\rm eq}|X_0)
+
Q_{\mathcal S}(t_{\rm eq}|X_0)
-
II(t_{\rm eq}|X_0).
\end{equation}
Substitution into Eqs.~\eqref{eq_fix_cmin} and \eqref{eq_fix_c_plu} yields
\begin{align}
\label{}
e^{-\beta\Delta F_{\mathcal S}^*}
&=
\frac{
\left\langle
e^{-\beta[W_{\mathcal S}+Q_{\mathcal S}-II]}
\right\rangle_{X_0}
}{
1+\chi^2(P_{t_{\rm eq}}^{\mathcal S}\parallel P_0^{\mathcal S})
},
\\
\label{}
e^{+\beta\Delta F_{\mathcal S}^*}
&=
\left\langle
e^{+\beta[W_{\mathcal S}+Q_{\mathcal S}-II]}
\right\rangle_{X_0}.
\end{align}

\section{Recovery of the Jarzynski equality}
\label{sec:SM_JE_limit}

\noindent
The HMF partition function satisfies
\begin{equation}
\label{}
\mathcal Z_{\mathcal S+\mathcal E}(\lambda,C,\beta)
=
\mathcal Z_{\mathcal S}^*(\lambda,C,\beta)\,
\mathcal Z_{\mathcal E}(\beta).
\end{equation}
Consequently,
\begin{align}
\label{eq:SM_partition_ratio}
e^{-\beta\Delta F_{\mathcal S}^*}
&=
\frac{
\mathcal Z_{\mathcal S}^*(\lambda(t_{\rm eq}),C(t_{\rm eq}),\beta)
}{
\mathcal Z_{\mathcal S}^*(\lambda(0),C(0),\beta)
}
=
\frac{
\mathcal Z_{\mathcal S+\mathcal E}(\lambda(t_{\rm eq}),C(t_{\rm eq}),\beta)
}{
\mathcal Z_{\mathcal S+\mathcal E}(\lambda(0),C(0),\beta)
}.
\end{align}
This identity becomes the Jarzynski equality only if the nonequilibrium dynamics allow the partition-function ratio to be written as an exponential work average.
In the standard deterministic Hamiltonian proof, this step relies on energy conservation along phase-space advection and on Liouville volume preservation \cite{jarzynski1997nonequilibrium,jarzynski1997equilibrium,jarzynski2004nonequilibrium}.
Analogous results can also be obtained for stochastic dynamics satisfying the usual canonical-invariance and local detailed-balance conditions.
Along a Hamiltonian trajectory,
\begin{align}
\frac{\mathrm d}{\mathrm dt}
\mathcal H_{\mathcal S+\mathcal E}(X(t),\lambda(t),C(t))
&=
\nabla_X\mathcal H_{\mathcal S+\mathcal E}\cdot\dot X
+
\frac{\partial\mathcal H_{\mathcal S}}{\partial\lambda}\dot\lambda
+
\frac{\partial\mathcal V_{\mathcal S\mathcal E}}{\partial C}\dot C.
\end{align}
The advection term vanishes because
\begin{equation}
\nabla_X\mathcal H_{\mathcal S+\mathcal E}\cdot\dot X
=
\{\mathcal H_{\mathcal S+\mathcal E},\mathcal H_{\mathcal S+\mathcal E}\}
=
0.
\end{equation}
Hence the parametric work on the composite is
\begin{equation}
\label{eq_work_compo}
W_{\rm comp}
=
\int_0^{t_{\rm eq}}\mathrm dt\,
\left[
\frac{\partial\mathcal H_{\mathcal S}}{\partial\lambda}\dot\lambda
+
\frac{\partial\mathcal V_{\mathcal S\mathcal E}}{\partial C}\dot C
\right],
\end{equation}
and
\begin{equation}
\mathcal H_{\mathcal S+\mathcal E}(X(t_{\rm eq}),\lambda(t_{\rm eq}),C(t_{\rm eq}))
-
\mathcal H_{\mathcal S+\mathcal E}(X_0,\lambda(0),C(0))
=
W_{\rm comp}.
\end{equation}
Using Liouville's theorem to change variables from $X_0$ to $X(t_{\rm eq})$,
\begin{align}
\left\langle e^{-\beta W_{\rm comp}}\right\rangle_{X_0}
&=
\frac{1}{\mathcal Z_{\mathcal S+\mathcal E}(\lambda(0),C(0),\beta)}
\int \mathrm dX_0\,
e^{-\beta\mathcal H_{\mathcal S+\mathcal E}(X(t_{\rm eq}),\lambda(t_{\rm eq}),C(t_{\rm eq}))}
\nonumber\\
&=
\frac{
\mathcal Z_{\mathcal S+\mathcal E}(\lambda(t_{\rm eq}),C(t_{\rm eq}),\beta)
}{
\mathcal Z_{\mathcal S+\mathcal E}(\lambda(0),C(0),\beta)
}.
\end{align}
Together with Eq.~\eqref{eq:SM_partition_ratio}, this gives
\begin{equation}
\label{eq_je_gen}
\left\langle e^{-\beta W_{\rm comp}}\right\rangle_{X_0}
=
e^{-\beta\Delta F_{\mathcal S}^*}.
\end{equation}
For frozen coupling, $C(t)\equiv C(0)$, Eq.~\eqref{eq_work_compo} reduces to the usual $\lambda$-work, and Eq.~\eqref{eq_je_gen} becomes the standard Jarzynski equality written in HMF notation.

\section{Scalar-action formulation}
\label{sec:SM_scalar_action}

\noindent
The identities can be reduced to a one-dimensional inference
problem.
This reduction is only statistical.
It does not mean that the
open-system dynamics become one-dimensional, nor that the many-particle
structure is neglected.
The microscopic system, the environment, the
system--environment coupling, and the trajectory history remain contained in
the sampled object.
What changes is only the level at which the identity is analyzed: instead of working directly with a full microscopic
configuration or a full trajectory, we work with the scalar quantity that
appears inside the exponential averages. 
Let
\begin{equation}
\label{}
Y=\mathcal Y(\Xi),
\end{equation}
where $\Xi$ denotes the sampled microscopic object and $\mu(\mathrm d\Xi)$ is
the sampling measure.
The symbol $\Xi$ is deliberately general.
It can
represent a final endpoint configuration, an initial condition propagated to
$t_{\rm eq}$, or an entire trajectory through the pathwise decomposition.
The
map $\mathcal Y$ assigns to each such microscopic object the scalar action
$Y$ that enters the relation.
For the fixed-coupling branch, three equivalent choices are useful:
\begin{align}
\Xi&=X_{\mathcal S},&
\mu(\mathrm d\Xi)&=
P_{t_{\rm eq}}^{\mathcal S}(X_{\mathcal S},\beta)\,\mathrm dX_{\mathcal S},
&
Y&=
\beta\Delta\mathcal H_{\mathcal S}(X_{\mathcal S}),
\label{eq_y_cho_end}
\\
\Xi&=X_0,&
\mu(\mathrm d\Xi)&=
P_0(X_0,\beta)\,\mathrm dX_0,
&
Y&=
\beta\Delta\mathcal H_{\mathcal S}(X_0),
\label{eq_y_cho_traj_end}
\\
\Xi&=X_0,&
\mu(\mathrm d\Xi)&=
P_0(X_0,\beta)\,\mathrm dX_0,
&
Y&=
\beta[W_{\mathcal S}+Q_{\mathcal S}-II].
\label{eq_cho_ac}
\end{align}
Equation~\eqref{eq_y_cho_end} uses final endpoint configurations
distributed according to the final HMF marginal.
Equation
\eqref{eq_y_cho_traj_end} uses initial conditions sampled from the
initial composite equilibrium and propagated to the final time.  Equation
\eqref{eq_cho_ac} uses the trajectory decomposition derived in
Sec.~\ref{sec:SM_heat_work}.
These three choices define equivalent scalar-action representations linked by
the exact identities.
The first representation is purely endpoint based.
The
second and third representations are trajectory based.
The probability density of $Y$ is the pushforward of the microscopic sampling
measure through the map $\mathcal Y$:
\begin{equation}
\label{}
p_Y(y)
=
\int \mu(\mathrm d\Xi)\,
\delta\!\left(y-\mathcal Y(\Xi)\right).
\end{equation}
This definition means that $p_Y(y)$ is obtained by taking all microscopic
samples $\Xi$, computing the scalar value $Y=\mathcal Y(\Xi)$ for each sample,
and forming the induced one-dimensional density.
The Dirac delta selects
those microscopic samples whose scalar value equals $y$. 
The pushforward definition has the defining property that every average over
$Y$ is equal to the corresponding microscopic average.
For any test function
$f$,
\begin{align}
\langle f(Y)\rangle
&=
\int \mu(\mathrm d\Xi)\,f(\mathcal Y(\Xi))
=
\int \mu(\mathrm d\Xi)
\int_{-\infty}^{\infty}\mathrm dy\,
\delta(y-\mathcal Y(\Xi))f(y)
=
\int_{-\infty}^{\infty}\mathrm dy\,
p_Y(y)f(y).
\label{}
\end{align}
Thus, no approximation is introduced by replacing the microscopic average by
an average over $p_Y$.
The approximation enters only later, when $p_Y$ is
reconstructed from finite samples.
With this
convention, the exact identities impose
\eqref{eq_fix_cmin} and \eqref{eq_fix_c_plu} become
\begin{align}
\label{eq_y_plu_ex}
\langle e^Y\rangle_Y
&=
e^{+\beta\Delta F_{\mathcal S}^*},
\\
\label{eq_y_min_ex}
\langle e^{-Y}\rangle_Y
&=
e^{-\beta\Delta F_{\mathcal S}^*}
\left[
1+\chi^2(P_{t_{\rm eq}}^{\mathcal S}\parallel P_0^{\mathcal S})
\right].
\end{align}
The positive exponential moment fixes the free energy displacement.  The
negative exponential moment contains the same free energy but also contains
the endpoint-overlap factor.
Multiplying the two moment constraints removes
$\Delta F_{\mathcal S}^{*}$ and gives
\begin{equation}
\label{eq_pro_exc_y}
\langle e^Y\rangle_Y\langle e^{-Y}\rangle_Y
=
1+\chi^2(P_{t_{\rm eq}}^{\mathcal S}\parallel P_0^{\mathcal S}).
\end{equation}
Equation~\eqref{eq_pro_exc_y} is the main reason for introducing the
scalar-action law.
It shows that the endpoint-overlap factor is encoded in the one-dimensional
random variable of $Y$.
In standard reweighting methods, overlap primarily appears as a numerical
limitation through variance growth and loss of effective sample size.
Here, the endpoint-overlap structure enters directly as a theoretical
quantity through the exact endpoint identities themselves.
The scalar-action formulation therefore converts overlap from a hidden
numerical difficulty into an explicit reconstructible observable.
Thus, the scalar density $p_Y$ contains two pieces of thermodynamic
information: the HMF free energy difference and the endpoint-overlap burden.

\section{Maximum-entropy reconstruction}
\label{sec:SM_MaxEnt}

\noindent
The exact constraints in Eqs.~\eqref{eq_y_plu_ex} and
\eqref{eq_y_min_ex} fix only two exponential moments of $p_Y$.
They
do not determine the full density.
Infinitely many distributions can share
the same values of $\langle e^Y\rangle_Y$ and $\langle e^{-Y}\rangle_Y$.
To obtain a unique scalar-action law without adding model-specific
assumptions, we choose the entropy maximizer compatible with these
constraints.
This is the maximum-entropy construction
\cite{jaynes1957information,shore2003axiomatic,cover1999elements}.
The maximum-entropy reconstruction is not introduced as an approximation to
the microscopic dynamics.
Rather, it is the least-structured scalar law compatible with the exact constraints. 
In the present setting, the maximum-entropy step has a precise role.  It does
not assume a Gaussian distribution for $Y$.
It also does not assume a
particular microscopic dynamics.
It only asks: among all one-dimensional laws
that reproduce the two exact exponential moments, which law adds the least
extra information?  The answer is the entropy maximizer.
In statistical
language, the resulting density belongs to an exponential family with
sufficient statistics $e^y$ and $e^{-y}$ \cite{wainwright2008graphical}. 
We maximize the differential entropy
\begin{equation}
\label{}
\mathbf S[p_Y]
=
-\int_{-\infty}^{\infty}\mathrm dy\,p_Y(y)\ln p_Y(y)
\end{equation}
subject to normalization,
\begin{equation}
\int_{-\infty}^{\infty}\mathrm dy\,p_Y(y)=1,
\end{equation}
and the two moment constraints,
\begin{equation}
\int_{-\infty}^{\infty}\mathrm dy\,e^y p_Y(y)=A_+,
\qquad
\int_{-\infty}^{\infty}\mathrm dy\,e^{-y}p_Y(y)=A_-,
\end{equation}
where
\begin{equation}
A_+=e^{+\beta\Delta F_{\mathcal S}^*},
\qquad
A_-=e^{-\beta\Delta F_{\mathcal S}^*}
\left[
1+\chi^2(P_{t_{\rm eq}}^{\mathcal S}\parallel P_0^{\mathcal S})
\right].
\end{equation}
The constants $A_+$ and $A_-$ are fixed by the exact identities.
They are not adjustable parameters in the formal derivation. 
Introduce Lagrange multipliers $\lambda_0$, $\lambda_+$, and $\lambda_-$.
The
constrained functional is
\begin{align}
\mathcal L[p_Y]
&=
-\int \mathrm dy\,p_Y\ln p_Y
-\lambda_0\left(\int \mathrm dy\,p_Y-1\right)
\nonumber\\
&\quad
-\lambda_+\left(\int \mathrm dy\,e^y p_Y-A_+\right)
-\lambda_-\left(\int \mathrm dy\,e^{-y}p_Y-A_-\right).
\label{}
\end{align}
The first term is the entropy.
The second term enforces normalization.
The
last two terms enforce the positive and negative exponential moments. 
The functional derivative of the entropy term is
\begin{equation}
\frac{\delta}{\delta p_Y(y)}
\left[
-\int \mathrm du\,p_Y(u)\ln p_Y(u)
\right]
=
-\ln p_Y(y)-1.
\end{equation}
This follows from the variation $p_Y\to p_Y+\varepsilon\eta$:
\begin{align}
\left.
\frac{\mathrm d}{\mathrm d\varepsilon}
\left[
-\int\mathrm dy\,
(p_Y+\varepsilon\eta)\ln(p_Y+\varepsilon\eta)
\right]
\right|_{\varepsilon=0}
&=
-\int \mathrm dy\,\eta(y)[\ln p_Y(y)+1].
\end{align}
The remaining derivatives are
\begin{align}
\frac{\delta}{\delta p_Y(y)}
\left[-\lambda_0\int\mathrm du\,p_Y(u)\right]
&=
-\lambda_0,
\\
\frac{\delta}{\delta p_Y(y)}
\left[-\lambda_+\int\mathrm du\,e^u p_Y(u)\right]
&=
-\lambda_+e^y,
\\
\frac{\delta}{\delta p_Y(y)}
\left[-\lambda_-\int\mathrm du\,e^{-u}p_Y(u)\right]
&=
-\lambda_-e^{-y}.
\end{align}
Stationarity of $\mathcal L[p_Y]$ gives
\begin{equation}
-\ln p_Y(y)-1-\lambda_0-\lambda_+e^y-\lambda_-e^{-y}=0.
\end{equation}
Solving for $p_Y$ gives
\begin{equation}
\label{eq:SM_pY_lambda}
p_Y(y)
=
\frac{1}{Z(\lambda_+,\lambda_-)}
\exp[-\lambda_+e^y-\lambda_-e^{-y}],
\end{equation}
where the normalization factor is
\begin{equation}
\label{}
Z(\lambda_+,\lambda_-)
=
\int_{-\infty}^{\infty}\mathrm dy\,
\exp[-\lambda_+e^y-\lambda_-e^{-y}].
\end{equation}
For normalizability, one needs $\lambda_+>0$ and $\lambda_->0$.
These
conditions ensure decay for $y\to+\infty$ and $y\to-\infty$, respectively. 
We now evaluate the normalization.
Set $t=e^y$, so that
$\mathrm dy=\mathrm dt/t$ and $t\in(0,\infty)$.
Then
\begin{equation}
Z(\lambda_+,\lambda_-)
=
\int_0^\infty\frac{\mathrm dt}{t}
\exp[-\lambda_+t-\lambda_-/t].
\end{equation}
Next define
\begin{equation}
u=\sqrt{\frac{\lambda_+}{\lambda_-}}\,t,
\qquad
\frac{\mathrm dt}{t}=\frac{\mathrm du}{u}.
\end{equation}
This gives
\begin{equation}
Z(\lambda_+,\lambda_-)
=
\int_0^\infty\frac{\mathrm du}{u}
\exp\left[
-\sqrt{\lambda_+\lambda_-}\left(u+\frac{1}{u}\right)
\right].
\end{equation}
Using the standard integral representation of the modified Bessel function of
the second kind \cite{gray1895treatise},
\begin{equation}
K_0(s)
=
\frac{1}{2}
\int_0^\infty\frac{\mathrm du}{u}
\exp\left[-\frac{s}{2}\left(u+\frac{1}{u}\right)\right],
\end{equation}
we find
\begin{equation}
Z(\lambda_+,\lambda_-)
=
2K_0\!\left(2\sqrt{\lambda_+\lambda_-}\right).
\end{equation}

\noindent
It is useful to reparametrize the two positive multipliers by
\begin{equation}
\label{eq_sm_def}
s=2\sqrt{\lambda_+\lambda_-}>0,
\qquad
m=\frac{1}{2}\ln\frac{\lambda_-}{\lambda_+}.
\end{equation}
The parameter $s$ controls the width and tail structure of the scalar-action
law.
Large $s$ gives a narrower distribution, while small $s$ gives broader
tails.
The parameter $m$ shifts the density along the $y$ axis and is tied to
the free energy differences.
Solving Eq.~\eqref{eq_sm_def} gives
\begin{equation}
\lambda_+=\frac{s}{2}e^{-m},
\qquad
\lambda_-=\frac{s}{2}e^{m}.
\end{equation}
Then
\begin{align}
\lambda_+e^y+\lambda_-e^{-y}
&=
\frac{s}{2}
\left(e^{y-m}+e^{-(y-m)}\right)
=
s\cosh(y-m).
\end{align}
Therefore, the least-biased scalar-action law is
\begin{equation}
\label{eq_prob_vari_exent}
p_Y(y;s,m)
=
\frac{1}{2K_0(s)}
\exp[-s\cosh(y-m)].
\end{equation}
This is the variational scalar-action law used in the main text.  It is the
maximum-entropy reconstruction of the scalar-action density under
normalization and the two exponential-moment constraints.  Its functional form
therefore does not depend on the microscopic model, the number of degrees of
freedom, or the chosen dynamics at the level of the variational
reconstruction.
All microscopic information enters through the sampled values
$Y_i$ and through the optimized parameters.
Equation~\eqref{eq_prob_vari_exent} reproduces the required moments while adding no model-specific structure beyond the imposed
constraints.

\section{Bessel moments and overlap relation}
\label{sec:SM_Bessel_moments}

\noindent
We now compute the exponential moments of Eq.~\eqref{eq_prob_vari_exent}.  This step
connects the maximum-entropy parameters $(s,m)$ to the physical quantities
$\Delta F_{\mathcal S}^{*}$ and
$1+\chi^2(P_{t_{\rm eq}}^{\mathcal S}\parallel P_0^{\mathcal S})$. 
For real $n$,
\begin{align}
\langle e^{nY}\rangle_Y
&=
\frac{1}{2K_0(s)}
\int_{-\infty}^{\infty}\mathrm dy\,
e^{ny}e^{-s\cosh(y-m)}.
\end{align}
With $u=y-m$, so that $y=u+m$, this becomes
\begin{align}
\langle e^{nY}\rangle_Y
&=
\frac{e^{nm}}{2K_0(s)}
\int_{-\infty}^{\infty}\mathrm du\,
e^{nu}e^{-s\cosh u}.
\end{align}
The modified Bessel function satisfies \cite{gray1895treatise}
\begin{equation}
K_n(s)
=
\frac{1}{2}
\int_{-\infty}^{\infty}\mathrm du\,
e^{-s\cosh u+nu}.
\end{equation}
Therefore,
\begin{equation}
\label{eq_mom_gene_var}
\langle e^{nY}\rangle_Y
=
e^{nm}\frac{K_n(s)}{K_0(s)}.
\end{equation}
This compact expression is the main reason why the Bessel parametrization is
useful.
All exponential moments are controlled by the ratio
$K_n(s)/K_0(s)$ and the shift factor $e^{nm}$. 
For $n=1$ and $n=-1$, Eq.~\eqref{eq_mom_gene_var} gives
\begin{align}
\label{eq_mom_plus_var_p}
\langle e^Y\rangle_Y
&=
e^m\frac{K_1(s)}{K_0(s)},
\\
\label{eq_mom_var_min_ps}
\langle e^{-Y}\rangle_Y
&=
e^{-m}\frac{K_1(s)}{K_0(s)},
\end{align}
where $K_{-1}(s)=K_1(s)$ was used.
Matching the positive moment
\eqref{eq_mom_plus_var_p} to the exact identity
\eqref{eq_y_plu_ex} gives
\begin{equation}
e^m\frac{K_1(s)}{K_0(s)}
=
e^{\beta\Delta F_{\mathcal S}^{*}}.
\end{equation}
Taking the logarithm yields
\begin{equation}
\label{}
\beta\Delta F_{\mathcal S}^*
=
m+\ln\frac{K_1(s)}{K_0(s)}.
\end{equation}
Equivalently,
\begin{equation}
\label{eq_m_sf_varit}
m=
\beta\Delta F_{\mathcal S}^*
-
\ln\frac{K_1(s)}{K_0(s)}.
\end{equation}
Thus, for a given shape parameter $s$, Eq.~\eqref{eq_m_sf_varit} determines the
location parameter $m$ from the free energy difference
$\Delta F_{\mathcal S}^{*}$.
The term $\ln[K_1(s)/K_0(s)]$ is the Bessel
correction that connects the statistical shape of $p_Y$ to the inferred free
energy difference.
Multiplying Eqs.~\eqref{eq_mom_plus_var_p} and
\eqref{eq_mom_var_min_ps} gives
\begin{equation}
\label{}
\langle e^Y\rangle_Y\langle e^{-Y}\rangle_Y
=
\left[\frac{K_1(s)}{K_0(s)}\right]^2.
\end{equation}
Comparison with the exact product identity
\eqref{eq_pro_exc_y} gives
\begin{equation}
\label{eq_chi_div_var}
\left[\frac{K_1(s)}{K_0(s)}\right]^2
=
1+\chi^2(P_{t_{\rm eq}}^{\mathcal S}\parallel P_0^{\mathcal S}).
\end{equation}
Equation~\eqref{eq_chi_div_var} is the overlap--Bessel relation.
It shows that the shape parameter $s$ carries the endpoint-overlap information.
Within the maximum-entropy scalar-action reconstruction, the endpoint-overlap
burden is therefore encoded in one scalar parameter.

\section{Likelihood estimator}
\label{sec:SM_likelihood}

\noindent
The scalar-action law in Eq.~\eqref{eq_prob_vari_exent} can be used as a variational
statistical model for sampled values
\begin{equation}
Y_i=\mathcal Y(\Xi_i),
\qquad
i=1,\ldots,N.
\end{equation}
Here each $Y_i$ is computed from one sampled microscopic object $\Xi_i$.
In
the validation, $\Xi_i$ denotes either a trajectory or an endpoint generated
from the simulation protocol, and $Y_i$ is the corresponding dimensionless
scalar action.
We then use the maximum-entropy scalar-action law as a
parametric density for the sampled values $\{Y_i\}_{i=1}^N$.
If the samples are independent replicas, the likelihood is the product of the
one-sample densities:
\begin{align}
L(s,m)
&=
\prod_{i=1}^N
\frac{1}{2K_0(s)}
\exp[-s\cosh(Y_i-m)]
=
[2K_0(s)]^{-N}
\exp\left[-s\sum_{i=1}^N\cosh(Y_i-m)\right].
\end{align}
The log-likelihood is
\begin{equation}
\label{eq_log_liki}
\ell(s,m)
=
-N\ln[2K_0(s)]
-
s\sum_{i=1}^N\cosh(Y_i-m).
\end{equation}
The maximum-likelihood principle chooses the parameters that make the observed
scalar-action sample most probable under the reconstructed law
\cite{fisher1922mathematical,cox2006principles}. 
The parameter $m$ is not the final physical quantity we want to report.
The
main physical target is $\Delta F_{\mathcal S}^{*}$.
We therefore eliminate
$m$ using Eq.~\eqref{eq_m_sf_varit}.
Define
\begin{equation}
\label{}
c(s)=\ln\frac{K_1(s)}{K_0(s)}.
\end{equation}
Then
\begin{equation}
m=\beta\Delta F_{\mathcal S}^*-c(s).
\end{equation}
Substitution into Eq.~\eqref{eq_log_liki} gives
\begin{align}
\label{}
\ell(s,\Delta F_{\mathcal S}^*)
&=
-N\ln[2K_0(s)]
-
s\sum_{i=1}^N
\cosh\left[
Y_i-\beta\Delta F_{\mathcal S}^*+c(s)
\right].
\end{align}
We minimize the negative log-likelihood,
\begin{align}
\label{eq_vari_J_minim}
J(s,\Delta F_{\mathcal S}^*)
&=
N\ln[2K_0(s)]
+
s\sum_{i=1}^N
\cosh\left[
Y_i-\beta\Delta F_{\mathcal S}^*+c(s)
\right].
\end{align}
The variational estimator is
\begin{equation}
\label{}
(\hat s,\widehat{\Delta F_{\mathcal S}^*})
=
\arg\min_{s>0,\;\Delta F_{\mathcal S}^*\in\mathbb R}
J(s,\Delta F_{\mathcal S}^*).
\end{equation}
The positivity constraint $s>0$ ensures that the distribution is normalizable.
In numerical applications, Eq.~\eqref{eq_vari_J_minim} is minimized as a variational
objective over the sampled scalar actions.
The minimizer gives the optimized
shape parameter and free energy within the maximum-entropy scalar-action
family.
For correlated samples, the Hessian uncertainty is interpreted as a
local sensitivity estimate.

\section{Profile reduction}
\label{sec:SM_profile}

\begin{figure*}[t]
\centering
\includegraphics[width=0.98\textwidth]{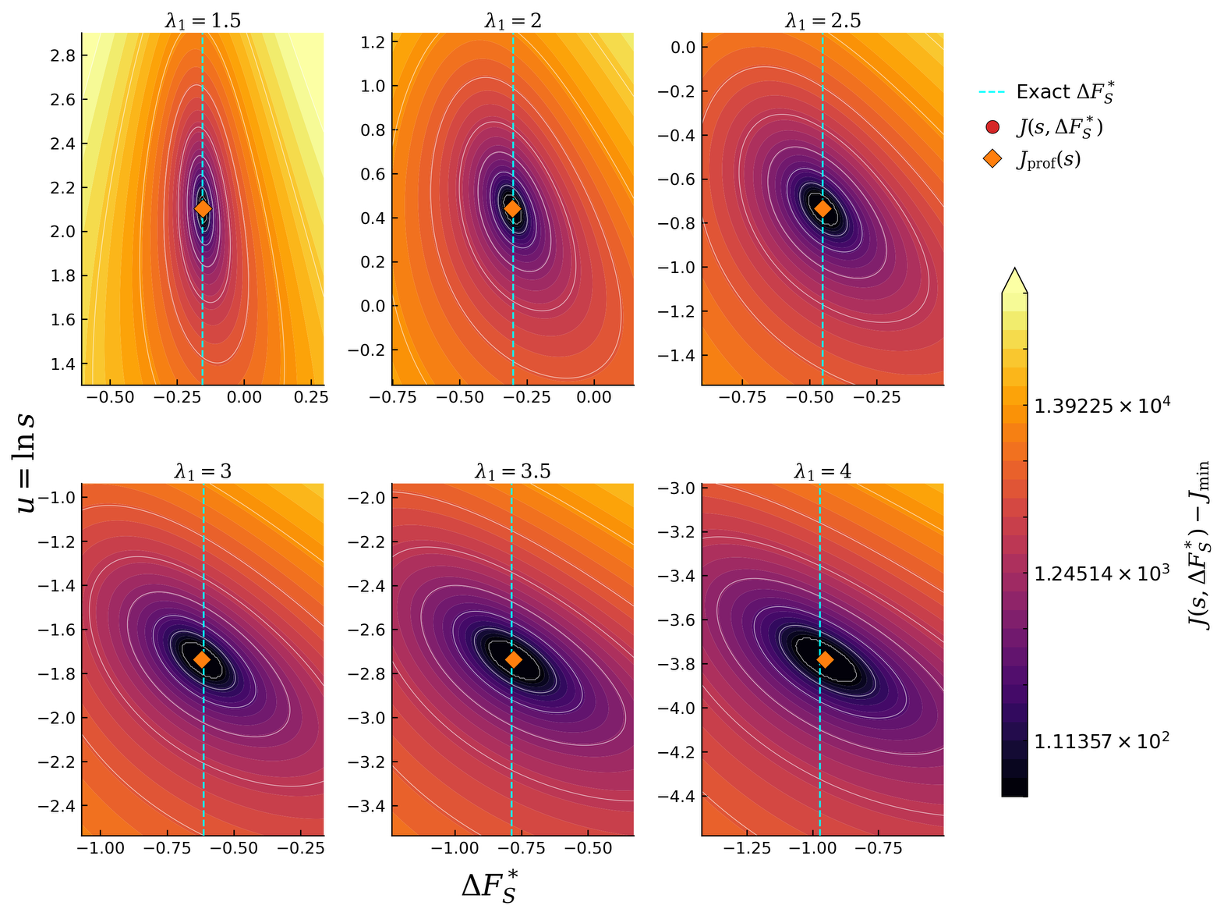}
\caption{
\textbf{Profile reduction of the variational objective.}
Each panel shows the shifted objective
$J(s,\Delta F_{\mathcal S}^{*})-J_{\min}$ in the
$(\Delta F_{\mathcal S}^{*},u)$ plane, with $u=\ln s$, for a different final
protocol value $\lambda_1=\lambda(t_{\rm eq})$.
The cyan dashed line marks
the exact HMF free energy differences.
The red circle denotes the minimum of the full
two-parameter objective $J(s,\Delta F_{\mathcal S}^{*})$, while the orange
diamond denotes the minimum obtained from the profile construction
$J_{\rm prof}(s)=J(s,\Delta F^*(s))$.
The near coincidence of the two markers
confirms the analytic profile reduction: for each fixed $s$, the minimizer in
$\Delta F_{\mathcal S}^{*}$ is unique and is given by
Eq.~\eqref{eq:SM_profile_F}.
The change in the contour geometry with
$\lambda_1$ reflects the changing endpoint-overlap burden encoded by the
optimized value of $s$.
}
\label{fig:SM_profile_landscapes}
\end{figure*}

\noindent
The two-parameter objective in Eq.~\eqref{eq_vari_J_minim} admits a useful reduction.
For fixed $s$, the minimization over $\Delta F_{\mathcal S}^{*}$ can be solved
analytically.
This profile construction leaves only the shape parameter $s$
to be optimized numerically.
Define
\begin{equation}
a_i(s)=Y_i+c(s).
\end{equation}
Then
\begin{equation}
J(s,\Delta F)
=
N\ln[2K_0(s)]
+
s\sum_{i=1}^N
\cosh[a_i(s)-\beta\Delta F].
\end{equation}
For fixed $s$, only the second term depends on $\Delta F$.  The first
derivative is
\begin{equation}
\label{}
\frac{\partial J}{\partial\Delta F}
=
-\beta s
\sum_{i=1}^N
\sinh[a_i(s)-\beta\Delta F],
\end{equation}
and the second derivative is
\begin{equation}
\label{}
\frac{\partial^2 J}{\partial(\Delta F)^2}
=
\beta^2 s
\sum_{i=1}^N
\cosh[a_i(s)-\beta\Delta F]>0.
\end{equation}
Since $s>0$ and $\cosh x>0$, the objective is strictly convex in
$\Delta F$ for every fixed $s$.
Therefore, the stationary point is the unique
minimum. 
The stationary condition is
\begin{equation}
\label{eq_station_pro}
\sum_{i=1}^N
\sinh[a_i(s)-\beta\Delta F]=0.
\end{equation}
Using
\begin{equation}
\sinh(x-y)=\sinh x\cosh y-\cosh x\sinh y,
\end{equation}
Eq.~\eqref{eq_station_pro} becomes
\begin{align}
0
&=
\cosh(\beta\Delta F)
\sum_{i=1}^N\sinh a_i(s)
-
\sinh(\beta\Delta F)
\sum_{i=1}^N\cosh a_i(s).
\end{align}
Rearranging gives
\begin{equation}
\tanh(\beta\Delta F)
=
\frac{
\sum_{i=1}^N\sinh a_i(s)
}{
\sum_{i=1}^N\cosh a_i(s)
}.
\end{equation}
The denominator is positive, and the ratio lies in $(-1,1)$ for finite data.
The unique profile minimizer is therefore
\begin{equation}
\label{eq:SM_profile_F}
\Delta F^*(s)
=
\frac{1}{\beta}
\operatorname{artanh}
\left[
\frac{
\sum_{i=1}^N\sinh[Y_i+c(s)]
}{
\sum_{i=1}^N\cosh[Y_i+c(s)]
}
\right].
\end{equation}
This expression gives the profiled estimate of $\Delta F_{\mathcal S}^{*}$ for
each prescribed shape parameter $s$. 
The one-dimensional profile objective is
\begin{equation}
\label{}
J_{\rm prof}(s)=J(s,\Delta F^*(s)).
\end{equation}
Thus,
\begin{equation}
\label{eq:SM_profile_estimator}
\hat s=\arg\min_{s>0}J_{\rm prof}(s),
\qquad
\widehat{\Delta F_{\mathcal S}^*}=\Delta F^*(\hat s).
\end{equation}
For numerical stability, one may use
\begin{equation}
s=e^u,
\qquad
u\in\mathbb R,
\end{equation}
and minimize $J_{\rm prof}(e^u)$.
The logarithmic variable removes the
positivity constraint on $s$ and makes standard unconstrained optimization
more stable. Fig.~\ref{fig:SM_profile_landscapes} verifies this profile reduction.

\section{Initialization and optimization schemes}
\label{sec:SM_optimization}

\noindent
The likelihood surface can be shallow in the
$(s,\Delta F_{\mathcal S}^{*})$ plane, especially for finite samples.  This
happens because the free energy differences and the shape parameter both
affect the argument
\begin{equation}
Y_i-\beta\Delta F_{\mathcal S}^{*}
+\ln\frac{K_1(s)}{K_0(s)}.
\end{equation}
A change in $s$ can therefore partly compensate a change in
$\Delta F_{\mathcal S}^{*}$.
Good initialization improves numerical
stability and reduces the chance that a local optimizer follows a long,
shallow valley before reaching the optimum.
A useful initial value for the free energy follows from the exact positive
moment in Eq.~\eqref{eq_y_plu_ex}:
\begin{equation}
\label{eq_initi_var}
\Delta F_{\rm init}
=
\frac{1}{\beta}
\ln\left[
\frac{1}{N}\sum_{i=1}^N e^{Y_i}
\right].
\end{equation}
This is the direct positive-exponential estimate.
It is not used as the final
answer.
It only provides a physically motivated starting value. 
A useful initial value for $s$ follows from the product relation
\eqref{eq_chi_div_var}.
Define the empirical product of exponential moments,
\begin{equation}
\label{}
M_{\rm init}
=
\left(
\frac{1}{N}\sum_{i=1}^N e^{Y_i}
\right)
\left(
\frac{1}{N}\sum_{i=1}^N e^{-Y_i}
\right).
\end{equation}
Then $s_{\rm init}$ is chosen as the solution of
\begin{equation}
\label{}
\left[\frac{K_1(s_{\rm init})}{K_0(s_{\rm init})}\right]^2
=
M_{\rm init}.
\end{equation}

\begin{figure}[t]
\centering
\includegraphics[width=0.98\columnwidth]{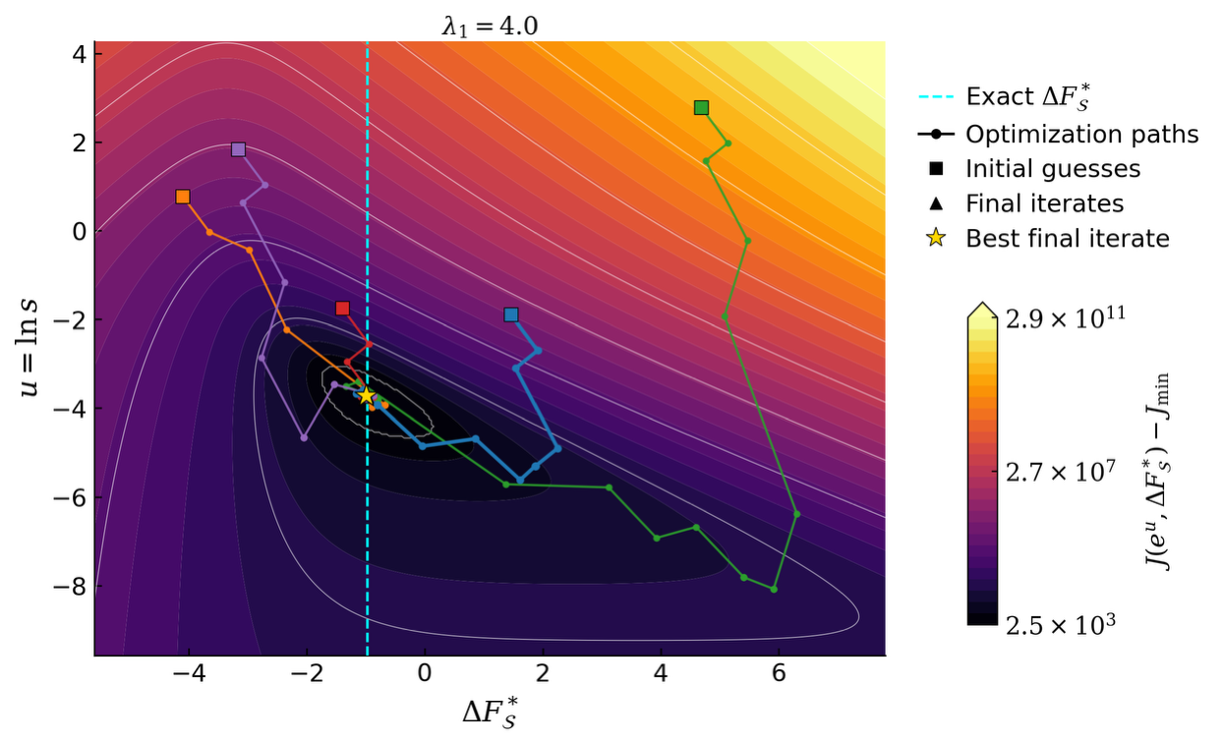}
\caption{
\textbf{Multistart optimization landscape for the variational estimator.}
The filled contours show the shifted objective
$J(e^u,\Delta F_{\mathcal S}^{*})-J_{\min}$ in the
$(\Delta F_{\mathcal S}^{*},u)$ plane, with $u=\ln s$, for the representative
protocol $\lambda(0)=1\to\lambda(t_{\rm eq})=4$.  The cyan dashed line marks
the exact HMF free energy differences $\Delta F_{\mathcal S}^{*}$.  Black squares denote
initial guesses, colored curves show representative Nelder--Mead optimization
paths, black triangles denote final iterates, and the yellow star marks the
best final iterate.
The broad valley illustrates why multistart optimization
is useful for the two-parameter optimization: the free energy differences and the
overlap-shape parameters are locally coupled, and poor initial guesses may
follow long curved paths before reaching the optimum.
}
\label{fig:SM_multistart_landscape}
\end{figure}

\noindent
Figure~\ref{fig:SM_multistart_landscape} shows the resulting optimization landscape.

\section{Hessian uncertainty}
\label{sec:SM_Hessian}

\begin{figure*}[t]
\centering
\includegraphics[width=\textwidth]{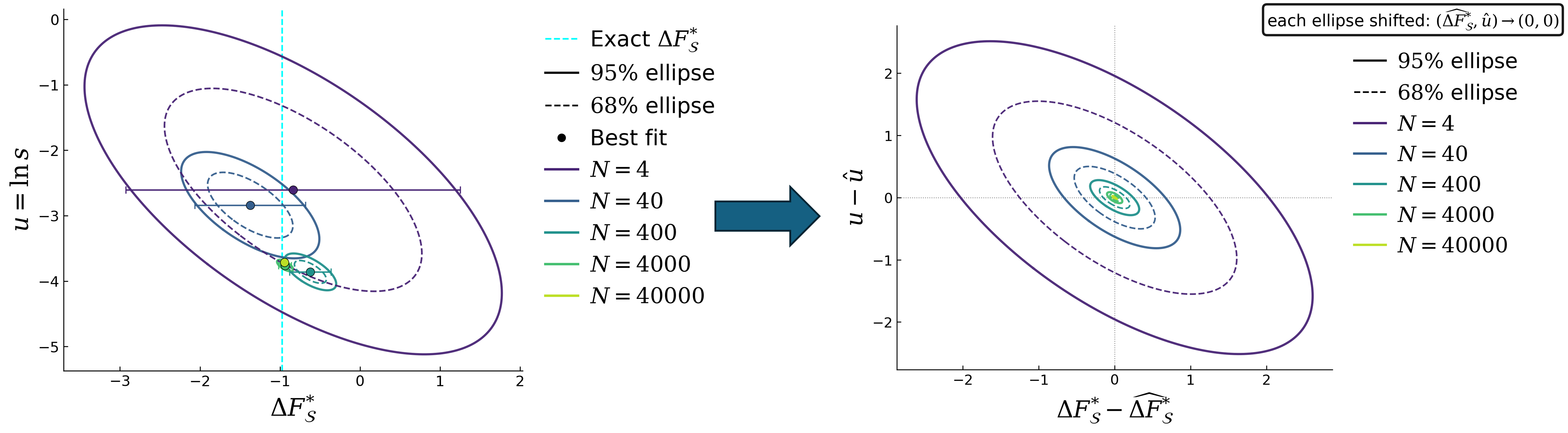}
\caption{
\textbf{Hessian uncertainty geometry for the variational estimator.}
Left: nominal joint local confidence ellipses in the
$(\Delta F_{\mathcal S}^{*},u)$ plane for increasing sample size $N$, with
$u=\ln s$.
Solid and dashed curves denote nominal $95\%$ and $68\%$ local
regions obtained from the Hessian approximation.
Filled circles show the
corresponding optima, and the cyan dashed line marks the exact HMF free
energy differences.
Horizontal bars show nominal one-parameter $95\%$ local intervals for
$\Delta F_{\mathcal S}^{*}$.
Right: the same ellipses shifted to their
individual optima, $(\widehat{\Delta F}_{\mathcal S}^{*},\hat u)\to(0,0)$.
This representation isolates the contraction and orientation of the local
covariance geometry.
The ellipses shrink with increasing sample size, showing
that the Hessian of $J_u$ gives a local uncertainty estimate for the free
energy differences and the overlap-shape parameter.
}
\label{fig:SM_hessian_ellipses}
\end{figure*}

\noindent
The variational optimization returns two coupled parameters: the free energy
$\Delta F_{\mathcal S}^{*}$ and the scalar-action shape parameter $s$.
Since
$s>0$, we optimize in the unconstrained variable
\begin{equation}
u=\ln s,
\qquad s=e^u .
\end{equation}
The parameter vector is
\begin{equation}
\boldsymbol\vartheta
=
\begin{pmatrix}
u\\
\Delta F_{\mathcal S}^{*}
\end{pmatrix}.
\end{equation}
In these variables, the negative log-likelihood is
\begin{align}
\label{}
J_u(u,\Delta F_{\mathcal S}^{*})
&=
N\ln[2K_0(e^u)]
+
e^u\sum_{i=1}^{N}
\cosh\!\left[
Y_i-\beta\Delta F_{\mathcal S}^{*}
+\ln\frac{K_1(e^u)}{K_0(e^u)}
\right].
\end{align}
This likelihood-based formulation follows the classical maximum-likelihood
principle, where parameters are obtained by minimizing the negative
log-likelihood \cite{fisher1922mathematical,cox2006principles}.
Related
likelihood and information-based constructions also appear in statistically
optimal free energy estimators such as BAR and MBAR
\cite{bennett1976efficient,shirts2008statistically}.
The optimum is denoted by
\begin{equation}
\hat{\boldsymbol\vartheta}
=
\begin{pmatrix}
\hat u\\
\widehat{\Delta F_{\mathcal S}^{*}}
\end{pmatrix},
\qquad
\hat s=e^{\hat u},
\end{equation}
and satisfies
\begin{equation}
\nabla_{\boldsymbol\vartheta}J_u
\big|_{\hat{\boldsymbol\vartheta}}
=
0
\end{equation}
up to numerical tolerance. 
The Hessian matrix is the matrix of second derivatives of $J_u$ at the
optimum,
\begin{equation}
\label{}
\mathbf H
=
\nabla_{\boldsymbol\vartheta}^{2}J_u
\big|_{\hat{\boldsymbol\vartheta}}
=
\begin{pmatrix}
H_{uu} & H_{u,\Delta F}\\[2pt]
H_{\Delta F,u} & H_{\Delta F,\Delta F}
\end{pmatrix}_{\hat{\boldsymbol\vartheta}} .
\end{equation}
For a smooth objective, $H_{u,\Delta F}=H_{\Delta F,u}$.
The Hessian measures
the local curvature of the objective.
Large curvature means that the data
strongly constrain the corresponding parameter, while small curvature signals
a shallow direction.
Thus, $\mathbf H$ acts as the observed-information matrix
of the variational optimization
\cite{fisher1922mathematical,cox2006principles,efron1978assessing}.
This
local information is also central in uncertainty estimates for free energy
methods, where curvature and overlap control estimator precision
\cite{bennett1976efficient,shirts2008statistically,klimovich2015guidelines,
pohorille2010good}. 
Near the optimum, the second-order expansion gives
\begin{equation}
\label{}
J_u(\boldsymbol\vartheta)
\simeq
J_u(\hat{\boldsymbol\vartheta})
+
\frac{1}{2}
(\boldsymbol\vartheta-\hat{\boldsymbol\vartheta})^{\rm T}
\mathbf H
(\boldsymbol\vartheta-\hat{\boldsymbol\vartheta}).
\end{equation}
Since the likelihood is proportional to $\exp[-J_u]$, this expansion gives the
local Gaussian approximation
\begin{align}
L(\boldsymbol\vartheta)
&\propto
\exp[-J_u(\boldsymbol\vartheta)]
\simeq
\exp[-J_u(\hat{\boldsymbol\vartheta})]
\exp\!\left[
-\frac{1}{2}
(\boldsymbol\vartheta-\hat{\boldsymbol\vartheta})^{\rm T}
\mathbf H
(\boldsymbol\vartheta-\hat{\boldsymbol\vartheta})
\right].
\end{align}
Under this approximation, the nominal local covariance matrix is
\begin{equation}
\label{eq_cova_va}
\boldsymbol\Sigma
\simeq
\mathbf H^{-1}.
\end{equation}
In component form,
\begin{equation}
\boldsymbol\Sigma
=
\begin{pmatrix}
\Sigma_{uu} & \Sigma_{u,\Delta F}\\[2pt]
\Sigma_{\Delta F,u} & \Sigma_{\Delta F,\Delta F}
\end{pmatrix}.
\end{equation}
The diagonal entries give the local uncertainties of $u$ and
$\Delta F_{\mathcal S}^{*}$, while the off-diagonal entry measures their local
correlation.
This correlation is expected because the objective contains the
coupled combination
\begin{equation}
Y_i-\beta\Delta F_{\mathcal S}^{*}
+\ln\frac{K_1(s)}{K_0(s)}.
\end{equation}
Changing $s$ changes the Bessel correction, while changing
$\Delta F_{\mathcal S}^{*}$ shifts the same argument in the opposite
direction. 
For the two-parameter Hessian
\begin{equation}
\mathbf H
=
\begin{pmatrix}
H_{uu} & H_{u,\Delta F}\\[2pt]
H_{u,\Delta F} & H_{\Delta F,\Delta F}
\end{pmatrix},
\end{equation}
the inverse is
\begin{equation}
\mathbf H^{-1}
=
\frac{1}{
H_{uu}H_{\Delta F,\Delta F}-H_{u,\Delta F}^{2}
}
\begin{pmatrix}
H_{\Delta F,\Delta F} & -H_{u,\Delta F}\\[2pt]
-H_{u,\Delta F} & H_{uu}
\end{pmatrix}.
\end{equation}
Since $\Delta F_{\mathcal S}^{*}$ is the second component of
$\boldsymbol\vartheta$, its local variance is the lower-right entry,
\begin{equation}
\label{eq_sigma_var_ex}
\Sigma_{\Delta F,\Delta F}
\simeq
\frac{
H_{uu}
}{
H_{uu}H_{\Delta F,\Delta F}-H_{u,\Delta F}^{2}
}.
\end{equation}
This expression accounts for the fact that $u=\ln s$ is inferred from the
same data.
If $u$ were fixed externally, the corresponding conditional
variance would be $1/H_{\Delta F,\Delta F}$.
The joint expression in
Eq.~\eqref{eq_sigma_var_ex} therefore includes the local coupling
between the shape parameter and the free energy estimate. 
The nominal local standard error of the free energy estimate is
\begin{equation}
\label{}
\sigma_{\Delta F}
=
\sqrt{\Sigma_{\Delta F,\Delta F}},
\end{equation}
or, equivalently,
\begin{equation}
\label{eq_sigma_var_ex_sqrt}
\sigma_{\Delta F}
\simeq
\left[
\frac{
H_{uu}
}{
H_{uu}H_{\Delta F,\Delta F}-H_{u,\Delta F}^{2}
}
\right]^{1/2}.
\end{equation}
A well-defined local covariance requires a positive-definite Hessian,
\begin{equation}
H_{uu}>0,
\qquad
\det\mathbf H
=
H_{uu}H_{\Delta F,\Delta F}-H_{u,\Delta F}^{2}
>0.
\end{equation}
If the determinant is small, the objective has a nearly flat direction and the
free energy uncertainty becomes large.
If the determinant is not positive,
the local quadratic approximation does not define a covariance matrix.
These
failures can occur under poor overlap, rare-event domination, or strongly
non-quadratic finite-sample objectives
\cite{gore2003bias,presse2006ordering,klimovich2015guidelines,
pohorille2010good}. 
A nominal one-parameter local interval for the free energy is obtained by
projecting the local Gaussian approximation onto the
$\Delta F_{\mathcal S}^{*}$ direction:
\begin{equation}
\label{eq:SM_CI_F}
\Delta F_{\mathcal S}^{*}
=
\widehat{\Delta F_{\mathcal S}^{*}}
\pm
z_{\alpha/2}\sigma_{\Delta F}.
\end{equation}
For the usual nominal $95\%$ local interval, $z_{0.025}=1.96$, giving
\begin{equation}
\Delta F_{\mathcal S}^{*}
=
\widehat{\Delta F_{\mathcal S}^{*}}
\pm
1.96\,\sigma_{\Delta F}.
\end{equation}
This interval keeps the uncertainty in $u$ through the inverse Hessian but
reports only the projected uncertainty along the free energy axis. 
The nominal joint local confidence region for both optimized parameters is
\begin{equation}
\label{}
(\boldsymbol\vartheta-\hat{\boldsymbol\vartheta})^{\rm T}
\mathbf H
(\boldsymbol\vartheta-\hat{\boldsymbol\vartheta})
\le
\chi^2_{2,p}.
\end{equation}
Here $\chi^2_{2,p}$ is the $p$ quantile of a chi-squared distribution with two
degrees of freedom.
This follows because the quadratic form
\begin{equation}
R
=
(\boldsymbol\vartheta-\hat{\boldsymbol\vartheta})^{\rm T}
\mathbf H
(\boldsymbol\vartheta-\hat{\boldsymbol\vartheta})
\end{equation}
is the squared Mahalanobis distance under the local Gaussian approximation.
For two degrees of freedom,
\begin{equation}
\Pr(R\le r)=1-e^{-r/2},
\end{equation}
so
\begin{equation}
\label{eq_chi_dive_quantile_explicit}
\chi^2_{2,p}
=
-2\ln(1-p).
\end{equation}
Thus,
\begin{equation}
\chi^2_{2,0.68}\simeq 2.28,
\qquad
\chi^2_{2,0.95}\simeq 5.99.
\end{equation}
These constants define the $68\%$ and $95\%$ ellipses.  They should not be
confused with the endpoint-overlap divergence
$\chi^2(P_{t_{\rm eq}}^{\mathcal S}\parallel P_0^{\mathcal S})$. 
Operationally, the ellipse is drawn from
\begin{equation}
\begin{pmatrix}
u-\hat u\\[2pt]
\Delta F_{\mathcal S}^{*}
-\widehat{\Delta F_{\mathcal S}^{*}}
\end{pmatrix}^{\rm T}
\mathbf H
\begin{pmatrix}
u-\hat u\\[2pt]
\Delta F_{\mathcal S}^{*}
-\widehat{\Delta F_{\mathcal S}^{*}}
\end{pmatrix}
=
\chi^2_{2,p}.
\end{equation}
Its center is the optimum, and its axes and orientation are fixed by the
eigenvectors and eigenvalues of $\mathbf H$.
A tilted ellipse indicates local
coupling between the inferred shape parameter and the free energy estimate.
This coupling is the geometric form of the same overlap sensitivity that
appears in free energy estimators through normalization, overlap, and sample
quality
\cite{bennett1976efficient,shirts2008statistically,klimovich2015guidelines}. 
The Hessian uncertainty is a local large-sample approximation within the
variational scalar-action model.
It is most reliable when the optimum is
interior, the objective is close to quadratic near the minimum, and the
scalar-action samples are effectively independent.
When samples are strongly
correlated or the objective is strongly non-quadratic, the inverse Hessian
should be read as a local sensitivity diagnostic rather than a strict
confidence interval.
This is the usual caution in finite-sample free energy
calculations, where poor overlap and rare events can make naive uncertainty
estimates unreliable
\cite{gore2003bias,presse2006ordering,klimovich2015guidelines,
pohorille2010good}. 
Fig.~\ref{fig:SM_hessian_ellipses} illustrates this construction.
For small sample size, the objective is shallow and the ellipse is broad.
As the sample size increases, the ellipse contracts around the optimum, the projected free energy interval narrows, and the optimized value approaches the exact HMF reference.
The right panel shifts all ellipses to their own optima, making the covariance contraction easier to compare.

\section{Validation model}
\label{sec:SM_model}

\noindent
The validation uses a one-dimensional double-well system bilinearly coupled to a harmonic environment.
Double-well models provide a standard test bed for activated dynamics and nonequilibrium free energy estimators \cite{hanggi1990reaction,whitelam2025improving}.
The system Hamiltonian is
\begin{equation}
\label{}
\mathcal H_{\mathcal S}(x,p_x;\lambda)
=
\frac{p_x^2}{2m}
+
U_{\mathcal S}(x;\lambda),
\end{equation}
with
\begin{equation}
\label{}
U_{\mathcal S}(x;\lambda)
=
\frac{1}{4}(x^2-\lambda)^2.
\end{equation}
The environment Hamiltonian is
\begin{equation}
\label{eq:SM_HE_model}
\mathcal H_{\mathcal E}(y,p_y)
=
\frac{p_y^2}{2m}
+
\frac{1}{2}\omega^2y^2,
\end{equation}
and the interaction is
\begin{equation}
\label{}
\mathcal V_{\mathcal S\mathcal E}(x,y;C)=Cxy.
\end{equation}
For fixed coupling, the HMF is
\begin{align}
\mathcal H_\beta^*(x,p_x;\lambda,C)
&=
\frac{p_x^2}{2m}
+
U_{\mathcal S}(x;\lambda)
-
\frac{C^2x^2}{2\omega^2}
+
\text{const.}
\label{eq_hmf_model}
\end{align}
The additive constant is independent of $x$, $p_x$, and $\lambda$ and cancels in normalized marginals and free energy differences at fixed $C$.  The result follows by completing the square in the environmental coordinate.  The configurational environmental factor is
\begin{align}
A(x,C)
&=
\frac{1}{Z_{\mathcal E}^{(y)}}
\int \mathrm dy\,
\exp\left[
-\beta\left(
\frac{1}{2}\omega^2y^2+Cxy
\right)
\right]
\nonumber\\
&=
\frac{1}{Z_{\mathcal E}^{(y)}}
\int \mathrm dy\,
\exp\left[
-\frac{\beta\omega^2}{2}
\left(y+\frac{Cx}{\omega^2}\right)^2
+
\frac{\beta C^2x^2}{2\omega^2}
\right]
\nonumber\\
&=
\exp\left(\frac{\beta C^2x^2}{2\omega^2}\right).
\end{align}
Thus,
\begin{equation}
e^{-\beta\mathcal H_\beta^*(x,p_x;\lambda,C)}
\propto
\exp\left[
-\beta\left(
\frac{p_x^2}{2m}
+
U_{\mathcal S}(x;\lambda)
-
\frac{C^2x^2}{2\omega^2}
\right)
\right],
\end{equation}
which gives Eq.~\eqref{eq_hmf_model}. 
The exact HMF reference is obtained from
\begin{equation}
\label{}
\mathcal Z_{\mathcal S}^*(\lambda,C,\beta)
=
\int \mathrm dx\,\mathrm dp_x\,
\exp\left[
-\beta\left(
\frac{p_x^2}{2m}
+
U_{\mathcal S}(x;\lambda)
-
\frac{C^2x^2}{2\omega^2}
\right)
\right].
\end{equation}
The free energy difference used in the validation is
\begin{equation}
\label{}
\Delta F_{\mathcal S,\rm exact}^*
=
-\frac{1}{\beta}
\ln
\frac{
\mathcal Z_{\mathcal S}^*(\lambda(t_{\rm eq}),C(0),\beta)
}{
\mathcal Z_{\mathcal S}^*(\lambda(0),C(0),\beta)
}.
\end{equation}

\section{Protocol and numerical details}
\label{sec:SM_protocol}

\noindent
The validation protocol has two stages.
During the ramp, the coupling is frozen,
\begin{equation}
C(t)\equiv C(0),
\end{equation}
and the control parameter changes linearly from $\lambda(0)$ to $\lambda(t_\lambda)$ over a time $t_{\rm ramp}$.
During this stage, the composite evolves under the deterministic non-Liouvillian equations
\begin{align}
\dot x&=\frac{p_x}{m},
\\
\dot p_x
&=
-\partial_xU_{\mathcal S}(x;\lambda(t))
-Cy
-\gamma_{\rm nl}^{\mathcal S}\frac{p_x}{m},
\\
\dot y&=\frac{p_y}{m},
\\
\dot p_y
&=
-\omega^2y-Cx
-\gamma_{\rm nl}^{\mathcal E}\frac{p_y}{m}.
\end{align}
The drag terms make the ramp phase-space compressing and break the Liouvillian assumption used in the standard Hamiltonian derivation of the Jarzynski equality.
This is the intended test.
The protocol violates the dynamical condition needed by the JE proof, while the final relaxation still prepares the correct canonical endpoint. 
After the ramp, $\lambda$ is held fixed and the composite relaxes under underdamped Langevin dynamics,
\begin{align}
\dot x&=\frac{p_x}{m},
\\
\dot p_x
&=
-\partial_xU_{\mathcal S}(x;\lambda(t_{\rm eq}))
-Cy
-\gamma\frac{p_x}{m}
+\eta_x(t),
\\
\dot y&=\frac{p_y}{m},
\\
\dot p_y
&=
-\omega^2y-Cx
-\gamma\frac{p_y}{m}
+\eta_y(t).
\end{align}
The noises are independent Gaussian white noises satisfying
\begin{equation}
\langle\eta_i(t)\rangle=0,
\qquad
\langle\eta_i(t)\eta_j(t')\rangle
=
\frac{2\gamma}{\beta}\delta_{ij}\delta(t-t'),
\end{equation}
for $i,j\in\{x,y\}$.
The Langevin stage obeys fluctuation--dissipation balance for the final frozen Hamiltonian and prepares the final canonical endpoint.
Thus it enforces the asymptotic equilibration condition used in Eq.~\eqref{eq_asymp}.  The relaxation stage is integrated with BAOAB, which is a standard splitting scheme for Langevin sampling \cite{kieninger2022gromacs}. 
The Jarzynski work estimator used for comparison is based on the ramp work
\begin{equation}
\label{}
W_\lambda
=
\int_0^{t_\lambda}\mathrm dt\,
\partial_\lambda U_{\mathcal S}(x_t;\lambda_t)\dot\lambda(t).
\end{equation}
For the non-Liouvillian ramp, the JE estimator
\begin{equation}
e^{-\beta\Delta F_{\mathcal S}^*}
\stackrel{\rm JE}{=}
\left\langle e^{-\beta W_\lambda}\right\rangle
\end{equation}
is not expected to hold, because the phase-space volume preservation step in the Hamiltonian proof is absent.
In contrast, Eqs.~\eqref{eq_fix_cmin} and \eqref{eq_fix_c_plu} use the final canonical endpoint and remain valid once the relaxation stage has prepared $P_{t_{\rm eq}}^{\mathcal S}$.

\begin{table}[t]
\centering
\caption{
Numerical inputs for the validation runs. All quantities are in reduced,
dimensionless units.
}
\label{tab:params}
\begin{ruledtabular}
\begin{tabular}{l l}
Quantity & Value \\
\hline
Mass $m$  & $1.0$ \\
System--environment coupling $C$ & $1.0$ \\
Inverse temperature $\beta$ & $1.0$ \\
Environment frequency $\omega$ & $1.0$ \\
Initial protocol $\lambda(0)$ & $1.0$ \\
Ramp duration $t_{\mathrm{ramp}}$ & $10.0$ \\
Relaxation duration $t_{\mathrm{eq}}-t_{\lambda}$ & $20.0$ \\
Ramp time step $\Delta t$ & $2\times10^{-3}$ \\
Relaxation time step $\Delta t_{\mathrm{relax}}$ & $2\times10^{-3}$ \\
Non-Liouvillian drag $\gamma_{\mathrm{nl}}^{\mathcal S}$ & $20.0$ \\
Non-Liouvillian drag $\gamma_{\mathrm{nl}}^{\mathcal E}$ & $5.0$ \\
Langevin damping $\gamma$ & $2.0$ \\
Initial ensemble size $N_{\mathrm{traj}}$ & $50000$ \\
Relaxation integrator & BAOAB \cite{kieninger2022gromacs} \\
\end{tabular}
\end{ruledtabular}
\end{table}

\section{Scalar-action samples used in the variational optimization}
\label{sec:SM_samples}

\noindent
For the fixed-coupling validation, the scalar action used in the variational construction is
\begin{equation}
\label{}
Y_i
=
\beta
\left[
\mathcal H_{\mathcal S}
(X_{\mathcal S}^{(i)}(t_{\rm eq}),\lambda(t_{\rm eq}))
-
\mathcal H_{\mathcal S}
(X_{\mathcal S}^{(i)}(t_{\rm eq}),\lambda(0))
\right].
\end{equation}
This is the dimensionless bare-system increment at the final system coordinate, evaluated between the final and initial values of the protocol.
Equivalently, when the pathwise decomposition is evaluated,
\begin{equation}
Y_i
=
\beta
\left[
W_{\mathcal S}^{(i)}
+
Q_{\mathcal S}^{(i)}
-
II^{(i)}
\right].
\end{equation}
The same sample $\{Y_i\}_{i=1}^N$ enters the variational density in
Eq.~\eqref{eq_prob_vari_exent}, the negative log-likelihood in Eq.~\eqref{eq_vari_J_minim},
the model-implied overlap reconstruction in Eq.~\eqref{eq_chi_div_var}, and the
Hessian uncertainty estimate in Eq.~\eqref{eq_cova_va}.  This is the main
practical advantage of the scalar-action formulation: one measured probability
distribution supplies the free energy estimate, an overlap diagnostic, and the
local uncertainty geometry.
For the representative protocol $\lambda(0)=1\to\lambda(t_{\rm eq})=4$, the optimized estimate reported in the main text is
\begin{equation}
\widehat{\Delta F}_{\mathcal S}^{*}=-0.9954,
\end{equation}
while the exact HMF reference is
\begin{equation}
\Delta F_{\mathcal S,\rm exact}^{*}=-0.9723.
\end{equation}
The nominal Hessian standard error is
\begin{equation}
\sigma_{\Delta F}=1.29\times10^{-2},
\end{equation}
which gives the nominal local $95\%$ interval
\begin{equation}
[-1.0207,-0.9700].
\end{equation}
The exact reference lies within this nominal local interval. 
The optimization-path, profile-landscape, and Hessian-ellipse diagnostics are shown in Figs.~\ref{fig:SM_multistart_landscape}, \ref{fig:SM_profile_landscapes}, and \ref{fig:SM_hessian_ellipses}.
Together, these figures document the numerical structure of the variational inference: the objective has a broad but well-defined valley, the analytic profile reduction reproduces the full two-parameter optimum, and the Hessian covariance contracts with increasing sample size.

\section{Liouvillian consistency check}
\label{sec:SM_liouvillian_check}

\noindent
As a consistency check, the non-Liouvillian drag terms during the ramp can be removed by setting
\begin{equation}
\gamma_{\rm nl}^{\mathcal S}=\gamma_{\rm nl}^{\mathcal E}=0.
\end{equation}
The ramp dynamics then become Hamiltonian for the composite system during the driving stage.
The JE proof in Sec.~\ref{sec:SM_JE_limit} applies, and the work estimator, the endpoint identities, and the exact HMF partition-function reference coincide.
This check separates two points.
First, when Liouville preservation holds, the standard JE is recovered.
Second, when the ramp is phase-space compressing, JE work reweighting is no longer protected by its Hamiltonian proof, whereas the endpoint and trajectory HMF identities remain valid after final equilibration.

\section{Multistage identities and overlap-resolved reconstruction}
\label{sec:SM_multistage_fixedC_estimators}

\noindent
The main article develops the frozen-coupling branch, $C(t)\equiv C(0)$.
In this branch, the external protocol $\lambda(t)$ drives the system, while
the coupling to the environment remains fixed.
The difference lies in the starting point.
The present theory does not begin from a work-reweighting identity.
It begins from canonical HMF endpoint marginals.
The free energy difference follows from exact endpoint identities, and the
trajectory form follows once the driven ensemble relaxes to the final
canonical HMF marginal.
This route separates the equilibrium endpoint structure from the dynamical
conditions used in the standard JE proof. 
A multistage version follows by inserting intermediate values of $\lambda$.
This construction leaves the exact free energy unchanged, but rewrites a large
transformation as a sequence of local HMF endpoint comparisons.
Each local step carries its own free energy increment, overlap factor, and
scalar-action law.
The multistage formulation therefore has two roles.
It reconstructs the total HMF free energy, and it identifies the parts of the
chosen path that are statistically costly.

\subsection{Endpoint multistage identity}

\noindent
We introduce a ladder of driving protocols at fixed coupling,
\begin{equation}
C(t)\equiv C(0),
\qquad
\lambda_0<\lambda_1<\cdots<\lambda_K ,
\qquad
\lambda_0=\lambda(0),
\qquad
\lambda_K=\lambda(t_{\rm eq}) .
\label{eq:SM_lambda_ladder}
\end{equation}
For each ladder state, define
\begin{equation}
P_k^{\mathcal S}(X_{\mathcal S},\beta)
=
P^{\mathcal S}(X_{\mathcal S};\lambda_k,C(0),\beta),
\end{equation}
and
\begin{equation}
F_{\mathcal S,k}^{*}
=
F_{\mathcal S}^{*}(\lambda_k,C(0),\beta).
\end{equation}
The total HMF free energy difference is the telescopic sum
\begin{equation}
\Delta F_{\mathcal S}^{*}
=
F_{\mathcal S,K}^{*}-F_{\mathcal S,0}^{*}
=
\sum_{k=1}^{K}
\Delta F_{\mathcal S,k}^{*},
\label{eq:SM_DF_multistage_sum}
\end{equation}
with
\begin{equation}
\Delta F_{\mathcal S,k}^{*}
=
F_{\mathcal S,k}^{*}-F_{\mathcal S,k-1}^{*}.
\end{equation}

\noindent
Since $C$ is fixed, the environmental contribution to the HMF is independent
of $\lambda$ and cancels in each local HMF difference.
The local HMF shift therefore reduces to the bare-system increment
\begin{equation}
\Delta\mathcal H_{\mathcal S,k}(X_{\mathcal S})
=
\mathcal H_{\mathcal S}(X_{\mathcal S},\lambda_k)
-
\mathcal H_{\mathcal S}(X_{\mathcal S},\lambda_{k-1}).
\label{eq:SM_local_DH_fixedC}
\end{equation}
Applying the endpoint identity to the pair
$(\lambda_{k-1},\lambda_k)$ gives
\begin{align}
e^{-\beta\Delta F_{\mathcal S,k}^{*}}
&=
\frac{
\left\langle
e^{-\beta\Delta\mathcal H_{\mathcal S,k}(X_{\mathcal S})}
\right\rangle_{k}
}{
1+\chi_k^2
},
\label{eq:SM_local_endpoint_minus}
\\
e^{+\beta\Delta F_{\mathcal S,k}^{*}}
&=
\left\langle
e^{+\beta\Delta\mathcal H_{\mathcal S,k}(X_{\mathcal S})}
\right\rangle_{k}.
\label{eq:SM_local_endpoint_plus}
\end{align}
Here
\begin{equation}
\langle A\rangle_k
=
\int \mathrm dX_{\mathcal S}\,
A(X_{\mathcal S})P_k^{\mathcal S}(X_{\mathcal S},\beta),
\end{equation}
and the local overlap factor is
\begin{equation}
1+\chi_k^2
=
\int \mathrm dX_{\mathcal S}\,
\frac{
\left[P_k^{\mathcal S}(X_{\mathcal S},\beta)\right]^2
}{
P_{k-1}^{\mathcal S}(X_{\mathcal S},\beta)
}
=
1+\chi^2(P_k^{\mathcal S}\parallel P_{k-1}^{\mathcal S}).
\label{eq:SM_local_chi_fixedC}
\end{equation}
Multiplying Eqs.~\eqref{eq:SM_local_endpoint_minus} and
\eqref{eq:SM_local_endpoint_plus} yields
\begin{equation}
\left\langle
e^{+\beta\Delta\mathcal H_{\mathcal S,k}}
\right\rangle_{k}
\left\langle
e^{-\beta\Delta\mathcal H_{\mathcal S,k}}
\right\rangle_{k}
=
1+\chi_k^2 .
\label{eq:SM_local_product_fixedC}
\end{equation}
Equation~\eqref{eq:SM_local_product_fixedC} is the local overlap-resolved
fixed-coupling identity.
It shows that each step contains two coupled pieces of information: the local
HMF free energy increment and the deformation of the reduced endpoint
marginal. 
The accumulated overlap burden of the chosen ladder is
\begin{equation}
\mathcal C_K
=
\sum_{k=1}^{K}\ln(1+\chi_k^2).
\label{eq:SM_cumulative_chi_fixedC}
\end{equation}
The quantity $\mathcal C_K$ is path dependent.
It is not, in general, equal to
$\ln[1+\chi^2(P_K^{\mathcal S}\parallel P_0^{\mathcal S})]$.
Rather, it measures how the endpoint mismatch is distributed over the chosen
intermediate protocols.
Large local values of $\ln(1+\chi_k^2)$ identify neighboring HMF marginals with
weak overlap.
Those stages are the natural targets for refinement.

\subsection{Trajectory multistage identity}

\noindent
Each local endpoint identity has a trajectory representation.
For the step $\lambda_{k-1}\to\lambda_k$, initial conditions are sampled from
the composite canonical state at $(\lambda_{k-1},C(0))$,
\begin{equation}
X_{k-1,0}\sim P(X;\lambda_{k-1},C(0),\beta).
\end{equation}
They are propagated under the chosen local protocol and then relaxed at fixed
$(\lambda_k,C(0))$ until the system marginal is $P_k^{\mathcal S}$.
Let $\mathcal T_k^{\mathcal S}(X_{k-1,0})$ denote the system coordinate at the
equilibrated end of this local stage.
The local trajectory increment is
\begin{align}
\Delta\mathcal H_{\mathcal S,k}(X_{k-1,0})
&=
\mathcal H_{\mathcal S}
(\mathcal T_k^{\mathcal S}(X_{k-1,0}),\lambda_k)
-
\mathcal H_{\mathcal S}
(\mathcal T_k^{\mathcal S}(X_{k-1,0}),\lambda_{k-1}).
\label{eq:SM_local_traj_DH_fixedC}
\end{align}
The local trajectory identities are
\begin{align}
e^{-\beta\Delta F_{\mathcal S,k}^{*}}
&=
\frac{
\left\langle
e^{-\beta\Delta\mathcal H_{\mathcal S,k}(X_{k-1,0})}
\right\rangle_{k-1,0}
}{
1+\chi_k^2
},
\label{eq:SM_local_traj_minus}
\\
e^{+\beta\Delta F_{\mathcal S,k}^{*}}
&=
\left\langle
e^{+\beta\Delta\mathcal H_{\mathcal S,k}(X_{k-1,0})}
\right\rangle_{k-1,0}.
\label{eq:SM_local_traj_plus}
\end{align}
The average $\langle\bullet\rangle_{k-1,0}$ is taken over the local initial
canonical composite ensemble and over the realized trajectory ensemble. 
The local heat--work--reference decomposition follows from the same algebra as
in Sec.~\ref{sec:SM_heat_work}, with $\lambda(0)$ and $\lambda(t_{\rm eq})$
replaced by $\lambda_{k-1}$ and $\lambda_k$:
\begin{equation}
\Delta\mathcal H_{\mathcal S,k}(X_{k-1,0})
=
W_{\mathcal S,k}
+
Q_{\mathcal S,k}
-
II_k .
\label{eq:SM_local_WQI_fixedC}
\end{equation}
The local terms are
\begin{equation}
W_{\mathcal S,k}
=
\int_{0}^{t_k}\mathrm dt\,
\frac{\partial\mathcal H_{\mathcal S}}
{\partial\lambda}
(X_{\mathcal S}(t),\lambda_k(t))
\dot\lambda_k(t),
\end{equation}
\begin{equation}
Q_{\mathcal S,k}
=
\int_{0}^{t_k}\mathrm dt\,
\nabla_{X_{\mathcal S}}\mathcal H_{\mathcal S}
(X_{\mathcal S}(t),\lambda_k(t))
\cdot
\dot X_{\mathcal S}(t),
\end{equation}
and
\begin{equation}
II_k
=
\int_{0}^{t_k}\mathrm dt\,
\nabla_{X_{\mathcal S}}\mathcal H_{\mathcal S}
(X_{\mathcal S}(t),\lambda_{k-1})
\cdot
\dot X_{\mathcal S}(t).
\end{equation}
Substitution into Eqs.~\eqref{eq:SM_local_traj_minus} and
\eqref{eq:SM_local_traj_plus} gives
\begin{align}
e^{-\beta\Delta F_{\mathcal S,k}^{*}}
&=
\frac{
\left\langle
e^{-\beta[W_{\mathcal S,k}+Q_{\mathcal S,k}-II_k]}
\right\rangle_{k-1,0}
}{
1+\chi_k^2
},
\label{eq:SM_local_WQI_minus}
\\
e^{+\beta\Delta F_{\mathcal S,k}^{*}}
&=
\left\langle
e^{+\beta[W_{\mathcal S,k}+Q_{\mathcal S,k}-II_k]}
\right\rangle_{k-1,0}.
\label{eq:SM_local_WQI_plus}
\end{align}
This form preserves the trajectory content at each stage.
The driven segment may be Hamiltonian, dissipative, stochastic, or
phase-space compressing.
The only endpoint condition is that the held protocol prepares the canonical
HMF marginal $P_k^{\mathcal S}$ after the local drive.

\subsection{Multistage scalar-action reconstruction}

\noindent
Each local stage defines a scalar action.
In endpoint form,
\begin{equation}
Y_k
=
\beta\Delta\mathcal H_{\mathcal S,k}(X_{\mathcal S}),
\qquad
X_{\mathcal S}\sim P_k^{\mathcal S}.
\end{equation}
In trajectory form,
\begin{equation}
Y_k
=
\beta\Delta\mathcal H_{\mathcal S,k}(X_{k-1,0}),
\qquad
X_{k-1,0}\sim P(X;\lambda_{k-1},C(0),\beta).
\end{equation}
Using the pathwise decomposition, the same scalar action is
\begin{equation}
Y_k
=
\beta[W_{\mathcal S,k}+Q_{\mathcal S,k}-II_k].
\end{equation}
The exact local moment constraints are
\begin{align}
\langle e^{Y_k}\rangle_{Y_k}
&=
e^{+\beta\Delta F_{\mathcal S,k}^{*}},
\label{eq:SM_local_Y_plus}
\\
\langle e^{-Y_k}\rangle_{Y_k}
&=
e^{-\beta\Delta F_{\mathcal S,k}^{*}}
(1+\chi_k^2),
\label{eq:SM_local_Y_minus}
\end{align}
and hence
\begin{equation}
\langle e^{Y_k}\rangle_{Y_k}
\langle e^{-Y_k}\rangle_{Y_k}
=
1+\chi_k^2.
\label{eq:SM_local_Y_product}
\end{equation}

\noindent
The maximum-entropy reconstruction applies independently to each stage.
For the $k$th scalar action, the least-biased law compatible with
Eqs.~\eqref{eq:SM_local_Y_plus} and \eqref{eq:SM_local_Y_minus} is
\begin{equation}
p_{Y_k}(y;s_k,m_k)
=
\frac{1}{2K_0(s_k)}
\exp[-s_k\cosh(y-m_k)].
\label{eq:SM_local_Bessel_law}
\end{equation}
Its Bessel moments give
\begin{equation}
\beta\Delta F_{\mathcal S,k}^{*}
=
m_k+\ln\frac{K_1(s_k)}{K_0(s_k)},
\label{eq:SM_local_DF_Bessel}
\end{equation}
and
\begin{equation}
\left[
\frac{K_1(s_k)}{K_0(s_k)}
\right]^2
=
1+\chi_k^2.
\label{eq:SM_local_chi_Bessel}
\end{equation}
Thus one local scalar-action law yields the free energy increment, the local
overlap diagnostic, and the local shape parameter. 
Given sampled scalar actions $\{Y_{k,i}\}_{i=1}^{N_k}$, the local negative
log-likelihood is
\begin{align}
J_k(s_k,\Delta F_{\mathcal S,k}^{*})
&=
N_k\ln[2K_0(s_k)]
+
s_k\sum_{i=1}^{N_k}
\cosh\left[
Y_{k,i}
-\beta\Delta F_{\mathcal S,k}^{*}
+
\ln\frac{K_1(s_k)}{K_0(s_k)}
\right].
\label{eq:SM_local_likelihood}
\end{align}
The local estimator is
\begin{equation}
(\hat s_k,\widehat{\Delta F_{\mathcal S,k}^{*}})
=
\arg\min_{s_k>0,\;\Delta F_{\mathcal S,k}^{*}\in\mathbb R}
J_k(s_k,\Delta F_{\mathcal S,k}^{*}).
\end{equation}
The multistage variational free energy estimate is
\begin{equation}
\widehat{\Delta F_{\mathcal S}^{*}}_{\rm multi}
=
\sum_{k=1}^{K}
\widehat{\Delta F_{\mathcal S,k}^{*}},
\label{eq:SM_multistage_variational_DF}
\end{equation}
and the cumulative Bessel overlap diagnostic is
\begin{equation}
\widehat{\mathcal C}_{K,{\rm Bessel}}
=
\sum_{k=1}^{K}
\ln
\left[
\left(
\frac{K_1(\hat s_k)}{K_0(\hat s_k)}
\right)^2
\right].
\label{eq:SM_multistage_variational_chi}
\end{equation}
For independently sampled local stages, the variance of the total estimate is
approximated by
\begin{equation}
\sigma_{\Delta F,{\rm multi}}^2
\simeq
\sum_{k=1}^{K}
\sigma_{\Delta F,k}^2,
\label{eq:SM_multistage_variance}
\end{equation}
where $\sigma_{\Delta F,k}^2$ follows from the inverse Hessian of $J_k$ in the
variables $(u_k,\Delta F_{\mathcal S,k}^{*})$, with $u_k=\ln s_k$.
If neighboring stages reuse samples or trajectories, the corresponding
cross-covariances must be retained.

\subsection{Numerical implementation of the multistage comparison}

\begin{figure*}[t]
\centering
\includegraphics[width=\textwidth]{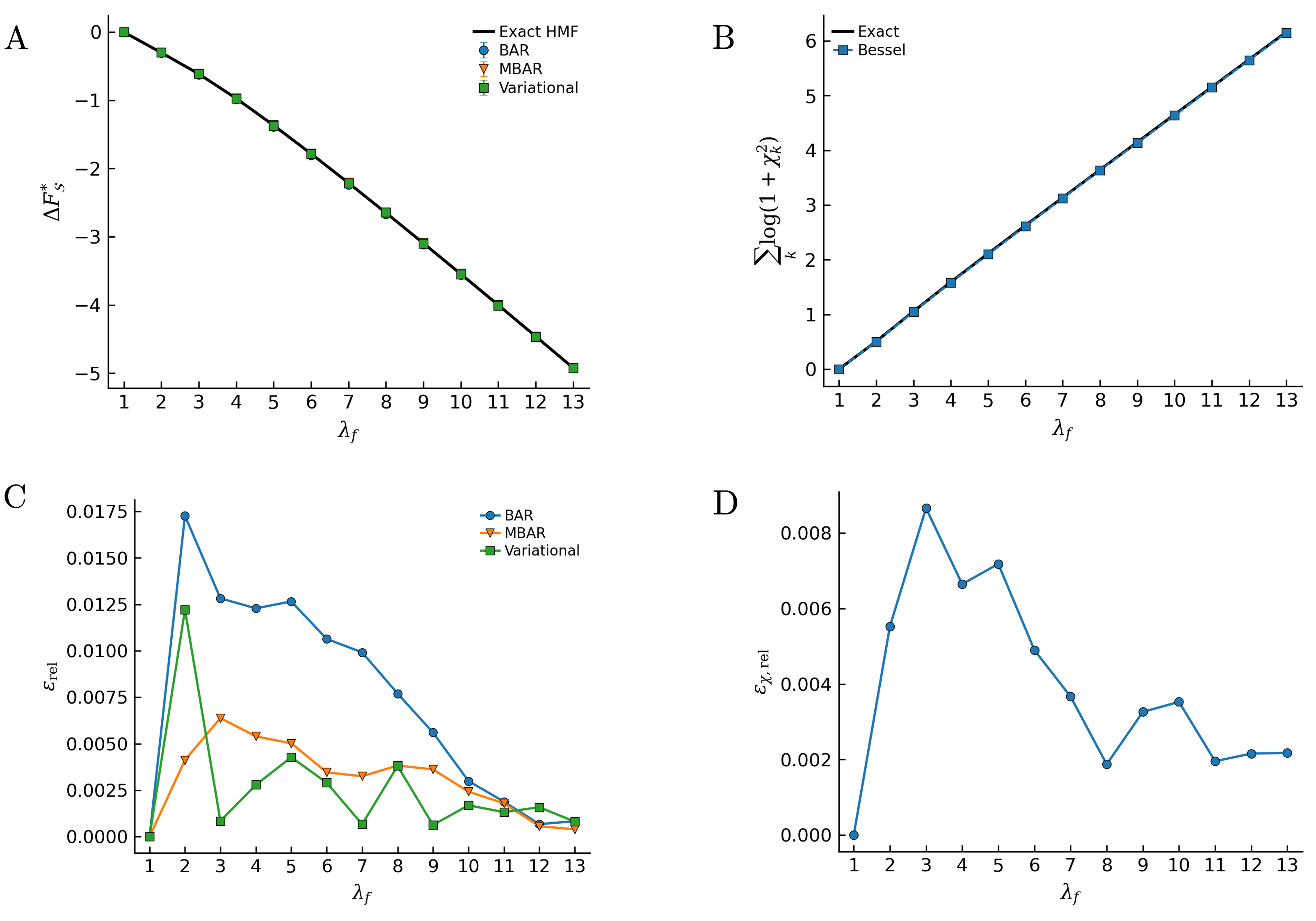}
\caption{
\textbf{Multistage frozen-coupling comparison.}
\textbf{(a)} Cumulative HMF free energy difference
$\Delta F_{\mathcal S}^{*}$ along the protocol ladder.
The solid line is the exact HMF partition-function reference; symbols show
BAR, MBAR, and the multistage variational scalar-action reconstruction.
\textbf{(b)} Cumulative overlap burden
$\sum_k\ln(1+\chi_k^2)$.
The exact value is compared with the Bessel reconstruction obtained from the
optimized scalar-action parameters.
\textbf{(c)} Relative free-energy error for BAR, MBAR, and the variational
method.
\textbf{(d)} Relative error of the Bessel reconstruction of the cumulative
overlap burden.
The protocol ladder starts at $\lambda_0=1$, and the error panels omit the
trivial reference point when appropriate.
}
\label{fig:SM_multistage_comparison}
\end{figure*}

\noindent
We applied the multistage construction to the frozen-coupling validation model
used in the main text.
The coupling was held fixed at $C=1$, and the driving ladder was
\begin{equation}
\lambda_0=1<\lambda_1<\cdots<\lambda_K=13,
\qquad
\lambda_k-\lambda_{k-1}=1 .
\label{eq:SM_lambda_ladder_numeric}
\end{equation}
Each local segment consists of a non-Liouvillian ramp from
$\lambda_{k-1}$ to $\lambda_k$, followed by underdamped Langevin relaxation at
fixed $\lambda_k$ and fixed $C$.
This prepares the final canonical HMF marginal $P_k^{\mathcal S}$ required in
Eqs.~\eqref{eq:SM_local_traj_minus} and \eqref{eq:SM_local_traj_plus}. 
The exact reference was obtained from the HMF partition functions,
\begin{equation}
\Delta F_{\mathcal S,{\rm exact}}^{*}(\lambda_j)
=
-\frac{1}{\beta}
\ln
\frac{
\mathcal Z_{\mathcal S}^{*}(\lambda_j,C,\beta)
}{
\mathcal Z_{\mathcal S}^{*}(\lambda_0,C,\beta)
}.
\label{eq:SM_exact_multistage_numeric}
\end{equation}
For the present model,
\begin{equation}
\mathcal H_\beta^*(x,p_x;\lambda,C)
=
\frac{p_x^2}{2}
+
\frac{1}{4}(x^2-\lambda)^2
-
\frac{C^2x^2}{2\omega^2},
\end{equation}
up to an additive constant independent of $\lambda$.
The quadrature of $\mathcal Z_{\mathcal S}^{*}$ therefore gives the exact HMF
curve shown in Fig.~\ref{fig:SM_multistage_comparison}. 
BAR and MBAR were evaluated as equilibrium baselines.
For BAR, independent composite equilibrium samples were generated at
neighboring states $\lambda_{k-1}$ and $\lambda_k$.
The local BAR increment was obtained from the usual two-state acceptance-ratio
equation and then accumulated over the ladder.
For MBAR, independent equilibrium samples from all ladder states were pooled,
and the multistate free energies were obtained from the standard MBAR
self-consistency equations.
Thus BAR and MBAR were tested in their natural equilibrium-sampling setting. 
For the variational reconstruction, the data at stage $k$ were the local
scalar actions
\begin{equation}
Y_{k,i}
=
\beta\left[
\mathcal H_{\mathcal S}
(\mathcal T_k^{\mathcal S}(X_{k-1,0}^{(i)}),\lambda_k)
-
\mathcal H_{\mathcal S}
(\mathcal T_k^{\mathcal S}(X_{k-1,0}^{(i)}),\lambda_{k-1})
\right].
\label{eq:SM_numeric_Yk}
\end{equation}
Equivalently, these are the sampled values of
$\beta(W_{\mathcal S,k}+Q_{\mathcal S,k}-II_k)$.
For each stage, $J_k$ in Eq.~\eqref{eq:SM_local_likelihood} was minimized over
$(s_k,\Delta F_{\mathcal S,k}^{*})$.
The local estimates were then summed according to
Eq.~\eqref{eq:SM_multistage_variational_DF}.
The same optimized parameters $\hat s_k$ also gave the Bessel estimate of the
local overlap factors through Eq.~\eqref{eq:SM_local_chi_Bessel}, and their
cumulative sum through Eq.~\eqref{eq:SM_multistage_variational_chi}. 
The four panels in Fig.~\ref{fig:SM_multistage_comparison} summarize the
calculation.
Panel (A) compares the cumulative free energy estimates with the exact HMF
reference.
BAR, MBAR, and the multistage variational reconstruction all follow the exact
curve over the full range $\lambda_f=1,\ldots,13$.
This agreement is expected for a sufficiently resolved ladder, where the
neighboring endpoint marginals remain well overlapped.
Panel (B) shows the cumulative overlap burden
$\sum_k\ln(1+\chi_k^2)$.
The Bessel estimate obtained from the variational scalar-action parameters is
indistinguishable from the exact value on the scale of the plot.
This confirms that the fitted scalar-action law reconstructs not only the
free energy but also the accumulated local overlap structure.
Panels (C) and (D) show the relative errors in the free energy and in the
cumulative overlap burden.
The errors remain small across the full ladder.
They also show the intended role of the construction: in this resolved regime,
the variational reconstruction gives the same HMF free energy accuracy as the
equilibrium baselines while additionally returning an explicit overlap
diagnostic from the same trajectory-generated scalar actions.

\subsection{Position relative to FEP, BAR, and MBAR}

\noindent
The multistage frozen-coupling construction places the present theory in the
same broad landscape as FEP, BAR, and MBAR, while addressing a different layer
of the free energy problem.
FEP, BAR, and MBAR are central equilibrium estimators for ratios of
normalization constants
\cite{zwanzig1954high,bennett1976efficient,shirts2008statistically}.
FEP uses a one-sided exponential average.
BAR combines the two endpoint ensembles and gives the optimal two-state
acceptance-ratio estimator in the large-sample limit.
MBAR extends this logic to several states.
These methods remain the appropriate statistical baselines when equilibrium
samples and reduced potentials are available. 
The present framework does not claim to supersede BAR or MBAR as equilibrium
estimators under their standard assumptions.
Its distinction is structural.
The primary objects are the reduced HMF marginals of an open system, and the
endpoint identities expose the overlap factor as an explicit term in the relation.
For each local step,
\begin{equation}
\left\langle e^{Y_k}\right\rangle
\left\langle e^{-Y_k}\right\rangle
=
1+\chi^2(P_k^{\mathcal S}\parallel P_{k-1}^{\mathcal S})
\end{equation}
links the scalar-action laws to the deformation of the reduced
endpoint marginal.
In standard reweighting, overlap controls estimator quality.
Here it also appears as a thermodynamic quantity fixed by the endpoint identities. 
This point also clarifies the relation to FEP.
The positive and negative exponential branches are algebraically related to
one-sided endpoint reweighting identities.
Taken separately, they do not remove the usual overlap limitation.
The new content lies in their joint use: their product gives the endpoint
overlap exactly, and the scalar-action reconstruction converts this overlap
factor into an inferred Bessel shape parameter.
Thus the method does not hide the overlap problem.
It reports it as part of the same inference problem that gives the HMF
free energy. 
The trajectory layer gives a second distinction.
BAR and MBAR do not depend on the dynamical route by which samples are
generated.
That is a strength when equilibrium samples from all states are available.
The present framework keeps the trajectory map in the formulation.
A local stage can be generated by Hamiltonian, dissipative, stochastic, or
phase-space-compressing dynamics, followed by relaxation to the final HMF
endpoint.
This is the regime in which the standard JE work average is not guaranteed by
Liouville preservation, while the endpoint and trajectory HMF identities remain
well defined after equilibration.
For each stage, the same scalar action can be written as an endpoint increment,
a trajectory increment, or a heat--work--reference action,
\begin{equation}
Y_k
=
\beta\Delta\mathcal H_{\mathcal S,k}
=
\beta(W_{\mathcal S,k}+Q_{\mathcal S,k}-II_k).
\end{equation}
The free energy calculation is therefore tied to a trajectory-generated scalar
observable, not only to equilibrium cross-evaluated weights. 
The likelihood structure is also different.
In BAR and MBAR, the estimating equations are built from equilibrium samples
assigned to thermodynamic states with known reduced potentials.
The data are the cross-evaluated reduced potentials $u_\ell(X_n)$, and the
unknowns are relative normalization constants.
The optimal weights are determined by how the sampled equilibrium ensembles
overlap in the full state space.
By contrast, the likelihood in Eq.~\eqref{eq:SM_local_likelihood} is built for
the one-dimensional pushforward law of the scalar action $Y_k$.
The data are the sampled values $\{Y_{k,i}\}$ generated by the
endpoint, trajectory, or heat--work--reference representation.
The fitted quantities are the local HMF free energy increment and the
scalar-action shape parameter.
Through the Bessel relation, the same shape parameter gives the local endpoint
overlap factor. 
The multistage setting gives this structure an operational role.
Standard stratification improves FEP, BAR, and MBAR by replacing a difficult
global transformation with better-overlapped local transformations.
The present theory gives the same strategy with an HMF-based interpretation: the free energy adds over stages, while the statistical cost is monitored by
\begin{equation}
\mathcal C_K
=
\sum_{k=1}^{K}\ln(1+\chi_k^2),
\end{equation}
or by its Bessel reconstruction
$\widehat{\mathcal C}_{K,{\rm Bessel}}$.
Large local contributions identify the parts of the chosen protocol path that
carry weak endpoint overlap and are natural targets for additional
intermediate states. 
The numerical comparison reflects this relation.
BAR and MBAR perform well, as they should, because the calculation supplies
equilibrium samples at each ladder state.
The multistage variational reconstruction reaches the same HMF free energy
accuracy in this resolved regime while using trajectory-generated
scalar-action samples and, at the same time, reconstructing the local overlap
burden.
The distinction is therefore not that equilibrium estimators such as BAR or
MBAR become invalid.
Rather, the present framework augments the free energy calculation with an
explicit overlap-resolved structure derived from reduced HMF
endpoint marginals. 
In this formulation, free energy estimation becomes an overlap-resolved inference problem rather than a pure reweighting problem.
This is the role of the multistage scalar-action construction: it connects
reduced endpoint marginals, trajectory-generated scalar actions, likelihood
uncertainty, and adaptive path refinement within one HMF framework.

\bibliography{apssamp}

\end{document}